\documentclass[sigconf]{aamas} 

\usepackage{balance} % for balancing columns on the final page
\usepackage{algorithm}
\usepackage{algorithmic}
\usepackage{amsfonts}
\usepackage{amsmath}

\usepackage{amssymb}
\usepackage{amsthm}
\usepackage{bm}
\usepackage{booktabs}
\usepackage{bxtexlogo}
\usepackage{caption}
\usepackage{graphicx} % DO NOT CHANGE THIS
\graphicspath{{figs/}}
\usepackage[utf8]{inputenc}
\usepackage{makecell}
\usepackage{mathtools}
\usepackage{multirow}
\usepackage{pifont}
\usepackage{soul}
\usepackage{subcaption}
\usepackage{xcolor}

\renewcommand\footnotetextcopyrightpermission[1]{} 
\title{Emergence from Emergence: Financial Market Simulation via Learning with Heterogeneous Preferences}

\author{Ryuji Hashimoto}
\affiliation{
  \institution{The University of Tokyo}
  \city{Tokyo}
  \country{Japan}}
\email{hashimoto-ryuji419@g.ecc.u-tokyo.ac.jp}

\author{Ryosuke Takata}
\affiliation{
  \institution{The University of Tokyo}
  \city{Tokyo}
  \country{Japan}}
\email{}

\author{Masahiro Suzuki}
\affiliation{
  \institution{The University of Tokyo}
  \city{Tokyo}
  \country{Japan}}
\email{}

\author{Yuki Tanaka}
\affiliation{
  \institution{The University of Tokyo}
  \city{Tokyo}
  \country{Japan}}
\email{}

\author{Kiyoshi Izumi}
\affiliation{
  \institution{The University of Tokyo}
  \city{Tokyo}
  \country{Japan}}
\email{}

\begin{abstract}
Agent-based models help explain stock price dynamics as emergent phenomena driven by interacting investors. In this modeling tradition, investor behavior has typically been captured by two distinct mechanisms---learning and heterogeneous preferences---which have been explored as separate paradigms in prior studies. However, the impact of their joint modeling on the resulting collective dynamics remains largely unexplored. We develop a multi-agent reinforcement learning framework in which agents endowed with heterogeneous risk aversion, time discounting, and information access collectively learn trading strategies within a unified shared-policy framework. The experiment reveals that (i) learning with heterogeneous preferences drives agents to develop strategies aligned with their individual traits, fostering behavioral differentiation and niche specialization within the market, and (ii) the interactions by the differentiated agents are essential for the emergence of realistic market dynamics such as fat-tailed price fluctuations and volatility clustering. This study presents a constructive paradigm for financial market modeling in which the joint design of heterogeneous preferences and learning mechanisms enables two-stage emergence: individual behavior and the collective market dynamics.
\end{abstract}


\newcommand{\BibTeX}{\rm B\kern-.05em{\sc i\kern-.025em b}\kern-.08em\TeX}

\begin{document}

%%% The following commands remove the headers in your paper. For final 
%%% papers, these will be inserted during the pagination process.

\pagestyle{fancy}
\fancyhead{}

%%% The next command prints the information defined in the preamble.

\maketitle 

%%%%%%%%%%%%%%%%%%%%%%%%%%%%%%%%%%%%%%%%%%%%%%%%%%%%%%%%%%%%%%%%%%%%%%%%

\section{Introduction}
% Agent-based models: constructive modeling, not analysing the real market, but constructing the artificial market by describing investor behavior. micro behavioral pattern to macro price dynamics
Agent-based models (ABMs)~\citep{abm_in_finance1,abm_in_finance2} simulate complex systems such as financial markets. ABMs in finance seek to construct artificial markets by designing investors' behavior as agents to understand how price dynamics emerge from their interactions~\citep{economy_needs_abm}.

% Two separate mechanisms driving agent behavior: Learning and Heterogeneity
% Gap: Joint modeling is unexplored. More descriptive, 2-staged emergence
To capture investors' behavior as agents, previous work explored two mechanisms---learning and heterogeneous preferences. Learning means that an investor discovers how to satisfy itself based on its own experiences~\citep{learning1,learning2,learning3,learning4}. Heterogeneous preferences mean differences in traits among investors~\citep{heterogeneous_preference2,heterogeneous_preference3}. Although recent studies~\citep{learning_and_heterogeneity1,learning_and_heterogeneity2,learning_and_heterogeneity3,adaptive_fcn_agent} introduced both mechanisms, the impact of such a joint modeling on the resulting dynamics remains unexplored.

% Methods
To investigate how learning and heterogeneous preferences jointly shape investor behavior and market dynamics, we employ a multi-agent reinforcement learning (MARL) framework. Within this framework, we design learning mechanism driven by multifaceted preferences. Agents are endowed with individual traits---including risk aversion, time discounting, and informational constraints---and approximate optimal shared-policy in a market environment through repeated interactions with one another.

% experiment: investigating RQs
In our experiment, we conduct simulations with the learned agents to investigate (i) behavioral differentiation arising from interactive learning within a unified framework where such heterogeneity in preferences exists and (ii) the emergence of collective market dynamics through the interactions of these agents. 
For (i), first, within a shared-policy environment, we analyze how agents’ trait values map to their order decisions, confirming that our unified learning mechanism produces behavioral differentiation rather than mere averaging across types. We conduct ablation studies that homogenize or scramble traits and find reduced social welfare, suggesting that the learned differentiation underpins niche specialization and a functionally complementary order.
For (ii), we show that our model achieves greater realism than baseline models that incorporate either preference heterogeneity or learning, based on several metrics. By combining market consensus learned through agent interactions with behavioral differentiation, key empirical regularities of financial markets emerge---such as fat-tailed return distributions and volatility clustering---that characterize the complex price dynamics observed in reality.

% Emergence of Emergence 
To summarize, through the joint modeling of learning and heterogeneous preferences, we enable {\em emergence from emergence}. This notion captures a hierarchical process of emergence. The first is behavioral differentiation at the individual level and niche specialization at the meso level. Within a unified learning framework, agents equipped with heterogeneous preferences competitively adapt in the same market environment. Despite sharing the same policy space, their strategies diverge endogenously, producing non-trivial diversity in investment behavior. Such differentiation fosters specialization into ecological niches and gives rise to a market-wide ecosystem of diverse trading roles. The second is the emergence of market-level dynamics. The interactions among these differentiated agents generate complex price dynamics that mirror the empirical regularities of real financial markets. Just as initially similar cells differentiate into distinct organs and self-organize into a life~\citep{collective_intelligence}, heterogeneous agents within a unified learning framework differentiate into diverse trading roles, which collectively self-organize into complex market dynamics.

\section{Related Work}
% 投資家の異質性が，金融市場のstylized factsや市場アノマリー，クラッシュなどの極端事象を再現する上で重要であるとの認識から，ABMによるアプローチが発達
% リスク回避度 -> FCN，
% time horizon -> ...
% (+ heterogeneous beliefを加える，．普通はchartistとfundamentalのweighting)
% 一方，市場のnonequibrium natureを再現するために学習メカニズムの設計も注目
The recognition that investor heterogeneity is crucial for reproducing stylized facts, market anomalies, and extreme events in financial markets has driven the development of ABM approaches~\citep{heterogeneous_preference2,heterogeneous_preference3}. For example, heterogeneity of risk aversions is proved to affect on equity risk premium~\citep{risk_aversion_term_to_risk_premium} and excess volatility~\citep{risk_aversion_term_to_excess_volatility1,risk_aversion_term_to_excess_volatility2}. \citet{dynamic_risk_aversion} demonstrate that heterogeneous and dynamic risk aversion of investors plays a role to generate volatility clustering. Also, heterogeneous time horizon is demonstrated to cause information delay, resulting in persistence of stock price fluctuations~\citep{time_horizon}.

To capture the non-equilibrium nature of financial markets, the design of learning mechanism has also attracted attention. A representative example is the adaptive belief system~\citep{learning2,learning3,adaptive_belief_system3}, a framework in which agents iteratively update their forecasting rules based on past performance, thereby allowing market expectations to evolve endogenously. Other approaches, such as classifier systems~\citep{learning4} and genetic algorithms~\citep{abm_in_finance1}, have likewise been employed to model how agents adapt to their environment and to explain the emergence of abrupt phenomena such as bubbles and crashes.

Although recent studies have incorporated both preference heterogeneity and learning into their models~\citep{learning_and_heterogeneity1,learning_and_heterogeneity2,learning_and_heterogeneity3} and have proposed methods to calibrate the hyperparameters representing heterogeneity~\citep{bounded_rational_rl,ot_based_simulation_evaluation}, the impact of jointly modeling these two factors on system complexity and the underlying emergent mechanisms remains largely unexplored. 

We argue that learning and preference heterogeneity are the key drivers for establishing a multi-level emergent view of financial markets. In biology, complex living systems are explained through hierarchical emergent mechanisms~\citep{collective_intelligence}, and similar multi-scale order is suggested to exist in human societies as another form of complex system~\citep{complexity_and_economy,complex_adaptive_system,future_economics}. However, a concrete explanation of such mechanisms in financial markets has not yet been attempted.

\section{Method}

\begin{figure}[tbp]
  \centering
  \includegraphics[width=\linewidth]{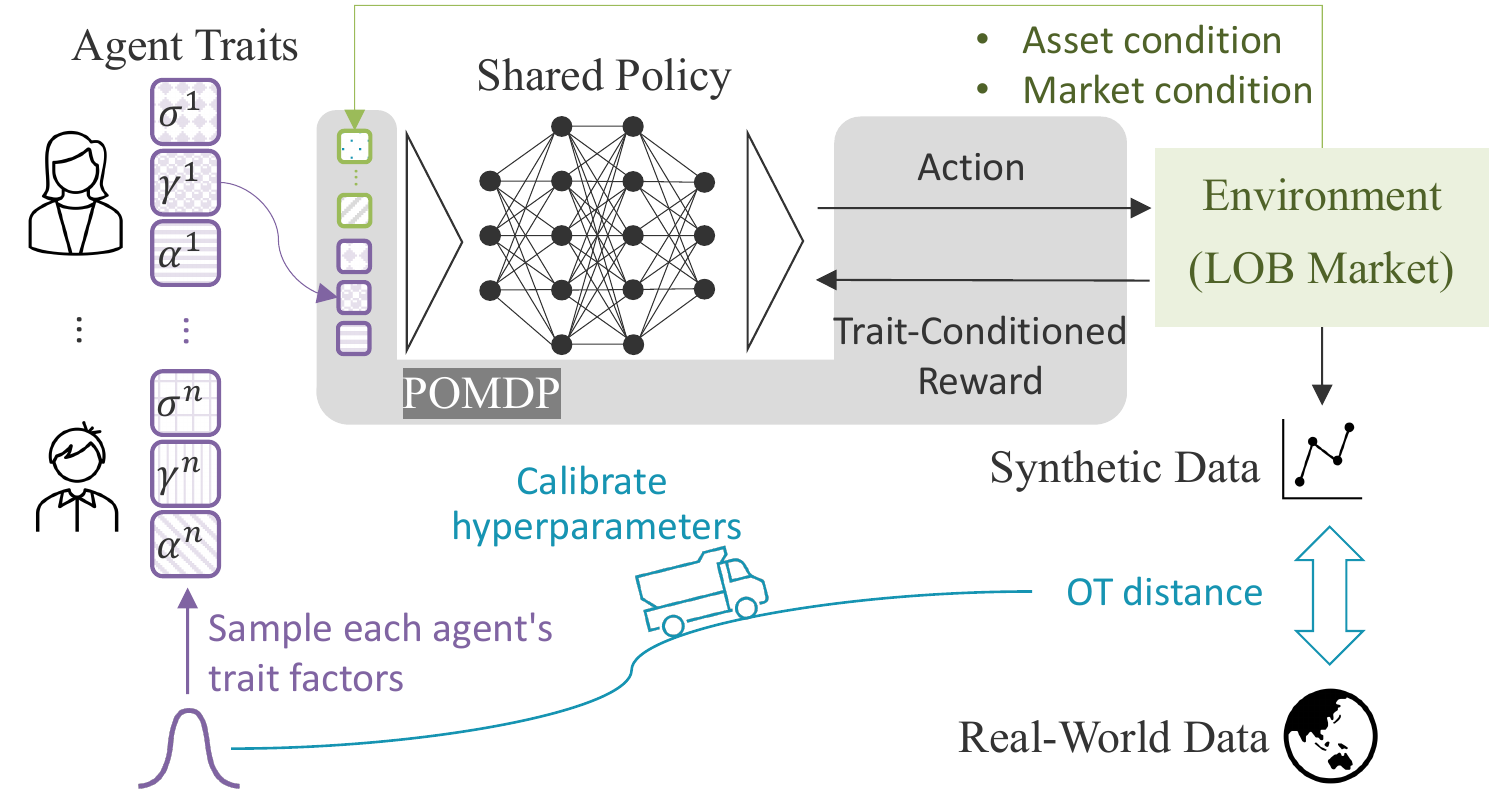}
  \caption{Conceptual diagram of our MARL-based ABM for financial market simulations. Each agent is assigned individual traits when the simulation starts. A shared-policy, learned to satisfy each agent's specific preference, governs agent behavior within a LOB market environment. The prior distribution of the agents' trait factors are calibrated using OT so that the synthetic price series aligns with real data.}
  \label{Fig:conceptual_diagram}
\end{figure}

Figure~\ref{Fig:conceptual_diagram} describes our simulation methodology. The core components of the method are a formulation as a partially observable Markov decision process (POMDP) in which agent traits are embedded in both the observations and the rewards, combined with shared-policy learning~\citep{shared_policy_learning,learning_and_heterogeneity3}, and a calibration of the distribution of agent traits via optimal transport (OT)~\citep{ot_based_simulation_evaluation}.
\begin{enumerate}
\item \textbf{POMDP formulation}: Each agent operates under a POMDP, where its observation vector includes trait factors, allowing the policy to condition on {\em who it is}. These traits are also reflected in the reward function, such that each agent is pursuing a different goal, encouraging the emergence of preference-driven behavioral diversity.
\item \textbf{Shared-policy learning across heterogeneous agents}: A single neural network policy is trained to map observations into actions, providing a unified learning framework within which heterogeneous traits give rise to endogenous behavioral differentiation.
\item \textbf{Data-driven calibration via OT}: The prior distribution of agent traits is calibrated at the population level to minimize the OT distance between the synthetic data generated by the simulation and real-world financial data, thereby enhancing the empirical realism of the model.
\end{enumerate}
This section describes the method in the following order: (1) the limit order book (LOB) market environment where agents interact; (2) the POMDP formulation defined by the observation, action, and reward structure; (3) the learning procedure of the shared-policy; and (4) the OT-based calibration of agent trait distributions.

\begin{figure}[tbp]
  \centering
  \includegraphics[width=0.98\linewidth]{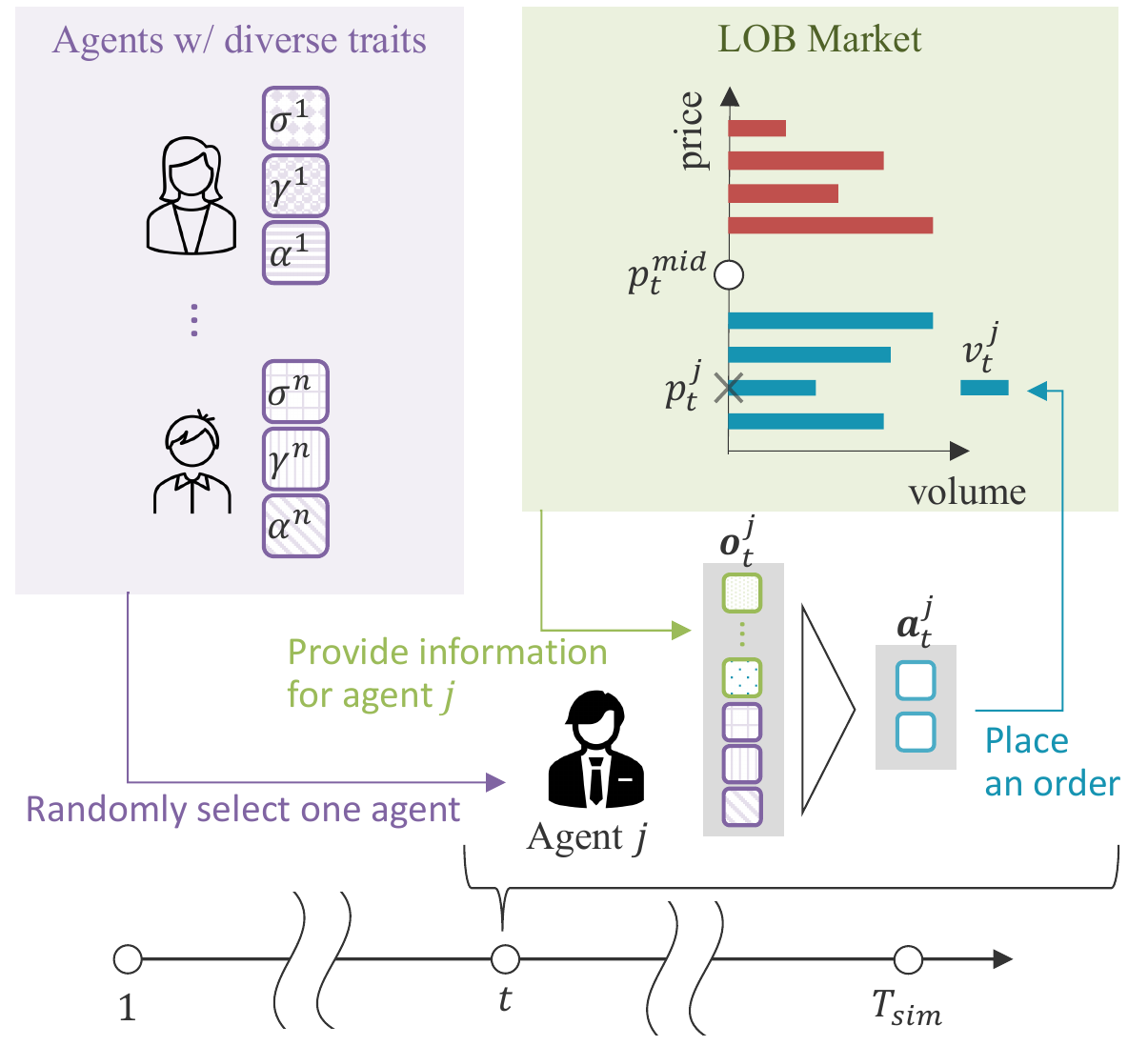}
  \caption{Structure of the simulation. At each time step $t$, one of the $n$ agents is selected to submit an order. Their order is placed into a LOB market, which matches buy and sell orders to determine transactions and update the market state.}
  \label{Fig:simulation_structure}
\end{figure}

\subsection{Simulation Structure}

Assume $n\in\mathbb{N}$ agents operate in a single market setting, with each simulation consisting of $T_{sim}\in\mathbb{N}$ time steps. Figure~\ref{Fig:simulation_structure} illustrates the structure of our simulation settings. At each time step $t \in \{1,\ldots,T_{sim}\}$, a randomly selected agent $j \in \{1,\ldots,n\}$ is allowed to place an order. Agent $j$ submits an order specifying a signed order volume $v_t^j \in \mathbb{Z}$ and an order price $p_t^j \in \mathbb{R}_+$. The sign of $v_t^j$ indicates whether the agent intends to buy or sell\footnote{For instance, if $v_t^j = -2$ and $p_t^j = 300.0$, agent $j$ submits an order to sell $2$ units of the stock at a price of 300.0.}.

The market operates under the double auction rule, where orders are collected into a central LOB. The LOB displays anonymized prices and volumes at which agents are willing to buy or sell the stock, and the market matches orders.

\subsection{POMDP and Shared-Policy Learning}
Let $\mathcal{T}^j$ denote the set of time steps at which agent $j$ is selected to place an order, with components $t_1^j < t_2^j < \ldots < t_{\iota_j}^j \in \mathcal{T}^j$, where $\iota_j$ represents the number of times agent $j$ is selected during the simulation. % satisfying the condition $\sum_j \iota_j = T_{sim}$. 
Agent $j$ aims to learn the optimal policy $\pi^{j*}$ that maximizes its expected cumulative discounted reward, defined as:
{\small
\begin{equation}
    \begin{split}
        \pi^{j*}=\arg\max_{\pi^j}\mathbb{E}_{\pi^j}\left[\sum_{i=1}^{\iota_j}(\gamma^j)^ig(\bm{o}_{t_i^j}^j,\bm{a}_{t_i^j}^j;j)\right]
    \end{split}
\end{equation}}
where $\mathbb{E}_{\pi^j}[\cdot]$ denotes the expectation under the policy $\pi^j$, and $\gamma^j$ is agent $j$'s discount factor. The reward function $g: \mathbb{R}^{11} \times \mathbb{R}^2 \to \mathbb{R}$ depends on the agent's observation $\bm{o}_{t_i^j}^j \in \mathbb{R}^{11}$ and action $\bm{a}_{t_i^j}^j \in \mathbb{R}^2$ at time step $t_i^j$.

\subsubsection{Observation}

Let $t = t_i^j$ in the following sections. The observation vector $\bm{o}_t^j \in \mathbb{R}^{11}$ is a numerical representation of the market conditions and agent $j$'s personal trait factors. It comprises the following eleven components:
\begin{itemize}
\item Holding asset ratio: $\frac{w_t^j p_t^{mid}}{x_t^j}$
\item Holding asset-to-maximum order volume ratio: $\frac{w_t^j}{v_{max}}$
\item Inverted buying power: $\frac{p_t^{mid}}{c_t^j}$
\item Return: $r_{[t_{i-1}^j, t_i^j]}$
\item Volatility: $V_{[t_{i-1}^j, t_i^j]}$
\item Asset volume-to-existing buy (sell) order volumes ratio: $\frac{|w_t^j|}{b_t^\xi},~\frac{|w_t^j|}{s_t^\xi}$
\item Blurred fundamental return: $\tilde{r}_t^{f,j}$
\item Uninformedness: $\sigma^j$
\item Risk aversion term: $\alpha^j$
\item Discount factor: $\gamma^j$
\end{itemize}
$w_t^j \in \mathbb{Z}$ and $c_t^j \in \mathbb{R}$ represent agent $j$'s stock position and holding cash amount at time step $t$, respectively. At the beginning of the simulation, $w_0^j$ and $c_0^j$ are randomly set as $w_0^j\sim Ex(w)$, $c_0^j\sim Ex(c)$ for each $j$, where $Ex(\lambda)$ indicates an exponential distribution with expected value $\lambda$, and $w$ and $c$ are the expected values of initial asset volume and cash amount. The agent's total asset value, $x_t^j$, is given by $x_t^j = c_t^j + w_t^j p_t^{mid}$, where $p_t^{mid}$ denotes the mid price at time $t$. \footnote{The mid price is calculated as the average of the best bid and ask prices. The best bid (ask) price refers to the highest (lowest) price buyers (sellers) are willing to trade on the LOB.}. $v_{max} \in \mathbb{N}$ is an exogenously determined constant that specifies the maximum order volume an agent can place at time step $t$. $r_{[t_{i-1}^j,t_i^j]} \in \mathbb{R}$ represents the logarithmic return of the price series between $t_{i-1}^j$ and $t_i^j$, calculated as follows:
{\small
\begin{equation}
    \begin{split}
    r_{[t_{i-1}^j,t_i^j]}=\frac{1}{t_i^j-t_{i-1}^j}\sum_{t'=t_{i-1}^j+1}^{t_i^j}\log\frac{p_{t'}^{mid}}{p_{{t'}-1}^{mid}}
    \end{split}
\end{equation}}
$V_{[t_{i-1}^j, t_i^j]}\in\mathbb{R}_+$ denotes the volatility:
{\small
\begin{equation}
    \begin{split}
    V_{[t_{i-1}^j, t_i^j]}=\frac{1}{t_i^j-t_{i-1}^j}\sum_{t'=t_{i-1}^j+1}^{t_i^j}\left(\log\frac{p_{t'}^{mid}}{p_{t'-1}^{mid}}-r_{[t_{i-1}^j,t_i^j]}\right)^2
    \end{split}
\end{equation}}
$b_t^\xi,s_t^\xi\in\mathbb{R}_+$ denote the weighted sum of order volumes within the price range on the LOB, extending up to $\xi p_t^{mid}$ away from the mid price $p_t^{mid}$
{\small
\begin{eqnarray}
b_t^\xi &=&\sum_{p_t^{mid}(1-\xi)<p<p_t^{mid}}v_t^b(p)\exp\left(-\omega^b\frac{|p_t^{mid}-p|}{p_t^{mid}}\right)\label{Eq:b_t_xi}\\
s_t^\xi &=&\sum_{p_t^{mid}<p<p_t^{mid}(1+\xi)}v_t^s(p)\exp\left(-\omega^s\frac{|p_t^{mid}-p|}{p_t^{mid}}\right)\label{Eq:s_t_xi}
\end{eqnarray}}
$\xi\in\mathbb{R}_+$ is a parameter that determines the maximum depth. $v_t^b(\cdot)$ and $v_t^s(\cdot)$ represent the buy and sell order volumes at specific prices displayed in the LOB. As shown in Equations~(\ref{Eq:b_t_xi}) and (\ref{Eq:s_t_xi}), the existing order volumes are weighted by their distances from the mid price. $\omega^b, \omega^s \in \mathbb{R}_+$ are parameters that control the decay rate. The agent $j$ receives a fundamental return blurred by Gaussian noise, the scale of which depends on their level of uninformedness $\sigma^j \in \mathbb{R}_+$.
{\small
\begin{equation}
    \begin{split}
    \tilde{r}_t^{f,j}=\log\frac{p_t^f}{p_t^{mid}}+\eta_t^j,~~\eta_t^j\sim\mathcal{N}(0,(\sigma^j)^2)\label{Eq:blurred_fundamental_return}
    \end{split}
\end{equation}}
$p_t^f$ represents the actual fundamental price of the stock. In real-world stock markets, investors estimate a company's intrinsic value, known as the fundamental price, based on various publicly available information and make trading decisions accordingly. An uninformedness $\sigma^j$ is introduced in Equation~(\ref{Eq:blurred_fundamental_return}) to capture differences in investors' ability to accurately perceive the fundamental price. The term $\mathcal{N}(0, (\sigma^j)^2)$ denotes a Gaussian distribution with a mean of $0$ and a variance of $(\sigma^j)^2$.

The agent $j$'s trait factors $\sigma^j$, $\alpha^j$, and $\gamma^j$ are randomly determined before each simulation as $\forall j~~\sigma^j\sim\mathcal{N}\left(\mu^\sigma,(\lambda^\sigma)^2\right),~\alpha^j\sim\mathcal{N}\left(\mu^\alpha,(\lambda^\alpha)^2\right),~\gamma^j\sim U(\lambda^\gamma,\gamma^{max})$ where $U(\lambda^\gamma,\gamma^{max})$ indicates uniform distribution with minimum and maximum values $\lambda^\gamma$ and $\gamma^{max}$.

\subsubsection{Action}
The action $\bm{a}_t^j\in\mathbb{R}^{2}$ consists of the following two components to form the agent $j$'s order.
\begin{itemize}
\item Scaled order volume $\tilde{v}_t^j$
\item Scaled order margin $\tilde{r}_t^j$
\end{itemize}
Using $\tilde{v}_t^j,\tilde{r}_t^j\in[-1,1]$, the signed order volume $v_t^j$ and order price $p_t^j$ is decided as follows.
{\small
\begin{equation}
    \begin{split}
    v_t^j=\lceil v_{max}\tilde{v}_t^j\rceil,~~ p_t^j=p_t^{mid}-r_{max}\cdot\mathrm{sign}(\tilde{v}_t^j)\cdot\tilde{r}_t^j\cdot p_t^{mid}\label{Eq:action2order}
    \end{split}
\end{equation}}
Here, $\lceil\cdot\rceil$ is a ceiling function. $r_{max}\in\mathbb{R}_+$ denotes the maximum scaled order price margin.

\subsubsection{Reward}

The immediate reward for agent $j$ at time step $t$ is defined as $r_t^j = g(\bm{o}_t^j, \bm{a}_t^j; j)$, where the reward function $g(\cdot, \cdot)$ is specified as follows.
{\small
\begin{equation}
    \begin{split}
    \MoveEqLeft g(\bm{o}_t^j,\bm{a}_t^j;j)=u_t^j-\beta^{short}\textbf{1}(w_t^j<0)-\beta^{cash}\textbf{1}(c_t^j<0)\\
    \MoveEqLeft~~~~~~~~~~~~~~~~~~~~~~-\beta^{illiquidity} l_t-\beta^{fundamental}R_t^f\label{Eq:reward}
    \end{split}
\end{equation}}
where, $\beta^{short}, \beta^{cash}, \beta^{illiquidity}, \beta^{fundamental} \in \mathbb{R}_+$ are weights corresponding to each component of the reward function. The components of the reward function (Equation~(\ref{Eq:reward})) can be categorized into terms related to individual learning (the first, second, and third terms) and those related to collective learning (the fourth and fifth terms). The first term, $u_t^j \in [-1,1]$, represents agent $j$’s utility: 
{\small
\begin{align}
\MoveEqLeft u_t^j=\frac{2}{\pi}\times\mathrm{Arctan}\nonumber\\
\MoveEqLeft \left(\omega^u(x_t^j+r_{[t_{i-1}^j,t_i^j]}w_t^jp_t^{mid}-\frac{\alpha^j}{2}V_{[t_{i-1}^j, t_i^j]}|w_t^jp_t^{mid}|)\right)\label{Eq:utility}
\end{align}}
where $\omega^u$ is a scaling factor for the utility. Equation~(\ref{Eq:utility}) models the agent $j$’s constant absolute risk averse (CARA) utility~\citep{fcn2}, capturing the decrease in utility due to volatility and the agent’s risk aversion parameter. The $\mathrm{Arctan}$ function compresses utility values to account for varying wealth scales. The second and third terms in Equation~(\ref{Eq:reward}) impose penalties for short selling and cash shortages, respectively, where $\textbf{1}(\cdot)$ denotes an indicator function. The fourth term represents an illiquidity penalty, reflecting investor discomfort when the stock market exhibits illiquidity. Illiquidity, $l_t$, is calculated as follows: 
{\small
\begin{equation}
    \begin{split}
    l_t=\frac{1}{b_t^\xi}+\frac{1}{s_t^\xi}+\omega^l\left(\frac{\max(b_t^\xi,s_t^\xi)}{\min(b_t^\xi,s_t^\xi)}-1\right)\label{Eq:illiquidity}
    \end{split}
\end{equation}}
where $l_t$ increases when either the buy or sell order volume in the LOB is low or when the order volumes are imbalanced. Here, $\omega^l$ is a scaling factor for the order volume imbalance. The fifth term in Equation~(\ref{Eq:reward}) is a penalty for deviation from the fundamental price. The integrated fundamental return, $R_t^f$, is calculated as follows:
{\small
\begin{align}
\begin{split}
\MoveEqLeft R_t^f = \sum_{\tau < t' \leq t} \left|\frac{1}{t+1-t'} \log \frac{p_{t'}^f}{p_{t'}^{mid}} \right| \\
\MoveEqLeft \text{s.t.}~ \mathrm{sign}\left( \log \frac{p_{\tau}^f}{p_{\tau}^{mid}} \right) \neq \mathrm{sign}\left( \log \frac{p_t^f}{p_t^{mid}} \right), \\
\MoveEqLeft ~\quad~ \mathrm{sign}\left( \log \frac{p_{t'}^f}{p_{t'}^{mid}} \right) = \mathrm{sign}\left( \log \frac{p_t^f}{p_t^{mid}} \right),~ \forall \tau < t' < t
\end{split}
\end{align}
}

% {\small
% \begin{align}
% \begin{split}
% \MoveEqLeft R_t^f=\sum_{\tau<t'\leq t}\left|\log\frac{p_{t'}^f}{p_{t'}^{mid}}\right|\\
% \MoveEqLeft s.t.~~\mathrm{sign}(\log\frac{p_{\tau}^f}{p_\tau^{mid}})\neq\mathrm{sign}(\log\frac{p_{t}^f}{p_t^{mid}}),\\
% \MoveEqLeft~~~~~~~~~ \forall\tau<t'~\mathrm{sign}(\log\frac{p_{t'}^f}{p_{t'}^{mid}})=\mathrm{sign}(\log\frac{p_{t}^f}{p_t^{mid}})
% \end{split}
% \end{align}}
The term $\beta^{fundamental} R_t^f$ imposes a larger penalty as the market price deviates from the fundamental price in the same direction for an extended period. This penalty encourage agents to collectively learn the consensus that market prices tend to fluctuate around the fundamental price.

Rather than hard-coding behavioral rules, we design reward functions that encode individual preferences. It specifies {\em what to optimize}, not {\em how to act}, providing a meta-level guidance for learning, without prescribing fixed action rules.

\subsubsection{Shared-Policy Learning}

\begin{algorithm}[t]
    \caption{The framework for learning heterogeneous trading policy in financial market simulations.}
    \label{Alg:learn_heterogeneous_marl}
    \begin{algorithmic}[1]
    \REQUIRE %$n$, the number of agents, $T_{sim}$, the number of total time steps in a simulation, $w$ and $c$, the expected values of initial asset volume and cash amount, $\lambda^\sigma$, $\lambda^{\gamma}$, and $\lambda^{\alpha}$, the parameters controlling the degree of heterogeneity in the agents' trait factors, $\mu^\sigma$, $\gamma^{max}$, and $\mu^{\alpha}$, the additional parameters for the prior distribution of the trait factors, $\omega^b$, $\omega^s$, $\omega^u$, and $\omega^l$ the scaling factors, $v_{max}$ and $r_{max}$, the maximum order volume and the scaled margin from the current mid price, $\beta^{short}$, $\beta^{cash}$, $\beta^{illiquidity}$, and $\beta^{fundamental}$, the weights of the reward function on the penalties, 
    $T_{rollout}$, the rollout length, $\beta^{actor}$ and $\beta^{critic}$, the learning rate for the actor and critic networks, %$\epsilon$, the clipping parameter for the surrogate objective of PPO, $\lambda$, discount factor for GAE, 
    and orthogonally initialized $\bm{\theta}$ and $\bm{\phi}$, the parameters for actor and critic networks, $L_{actor}(\cdot)$ and $L_{critic}(\cdot)$, loss functions for updating the networks. %$\pi_{\bm{\theta}}$ and $\bm{\phi}$, the parameters for critic network $C_{\bm{\phi}}$, $L_{actor}(\cdot)$ and $L_{critic}(\cdot)$, loss functions for updating actor and critic networks.
    \STATE Set rollout buffer for all agents empty $\forall j~ D_j=\{\}$
    \WHILE{not converge}
        \STATE Set trait factors of all agents by randomly sample $\forall j~w_0^j,c_0^j,\sigma^j,\gamma^j$, and $\alpha^j$ from prior distributions.%$\forall j~w_0^j\sim Ex(w)$, $c_0^j\sim Ex(c)$, $\sigma^j\sim\mathcal{N}(\mu^\sigma,\lambda^\sigma)$, $\gamma^j\sim U(\lambda^\gamma, \gamma^{max})$, and $\alpha^j\sim\mathcal{N}(\mu^\alpha,\lambda^\alpha)$
        \STATE Set previous order times of all agents $\forall j~ t_0^j\leftarrow0$
        \FOR{$t\in \{1,\ldots,T_{sim}\}$}
            \STATE Update fundamental price $p_t^f$
            \STATE Randomly sample an agent $j\in\{1,\ldots, n\}$
            \STATE Retrieve the last order time of agent $j$, $t_{i-1}^j$ and set $t_i^j \leftarrow t$ (current time step, $i\in\{1,\ldots\iota_j\}$)
            \STATE Get observation for the agent $j$, $\bm{o}_{t_i^j}^j$
            \STATE Sample action $\bm{a}_{t_i^j}^j$ and its probability {\small $\pi_{\bm{\theta}_{t_i^j}}(\bm{a}_{t_i^j}^j | \bm{o}_{t_i^j}^j)$}
            \STATE Convert action $\bm{a}_{t_i^j}^j$ to an order, submit the order to the market, and the market execute orders
            \STATE Calculate reward for agent $j$, $r_{t_i^j}^j = g(\bm{o}_{t_i^j}^j, \bm{a}_{t_i^j}^j; j)$
            \IF{$1<i$}
                \STATE {\small $D_j=D_j~\cup~\left[\bm{o}_{t_{i-1}^j}^j,\bm{a}_{t_{i-1}^j}^j,\log \pi_{\bm{\theta}_{t_{i-1}^j}}, r_{t_i^j}^j,\bm{o}_{t_i^j}^j\right]$}
                \IF{length of $D_j$ equals $T_{rollout}$}
                    \STATE Update actor and critic using rollout {\small $D_j$, $\bm{\theta}_t\leftarrow\bm{\theta}_{t-1}-\beta^{actor}\nabla L_{actor}(\bm{\theta}_{t-1})$, $\bm{\phi}_t\leftarrow \bm{\phi}_{t-1}-\beta^{critic}\nabla L_{critic}(\bm{\phi}_{t-1})$}
                    \STATE Clear rollout buffer for agent $j$ $D_j=\{\}$
                \ENDIF
            \ENDIF
        \ENDFOR
    \ENDWHILE
    \end{algorithmic}
\end{algorithm}

We train a single shared-policy across all agents using proximal policy optimization (PPO)~\citep{ppo}. The details of the learning procedure are outlined in Algorithm~\ref{Alg:learn_heterogeneous_marl}. To preserve individual-level variation and stabilize learning, each agent maintains its own experience buffer $D_j$, and the policy $\pi_{\bm{\theta}}$ is updated only when each buffer reaches a fixed rollout length $T_{rollout}$. This design is aimed to prevent the mixing of trajectories generated under different agent preferences, which could otherwise dilute the distinct behavioral strategies shaped by their respective traits.

%To capture heterogeneous behaviors while maintaining scalability, we train a single shared-policy across all agents using proximal policy optimization (PPO)~\citep{ppo}. The details of the learning procedure are outlined in Algorithm~\ref{Alg:learn_heterogeneous_marl}. To preserve individual-level behavioral variation and ensure learning stability, each agent maintains its own experience buffer $\forall j~~D_j$, and the shared-policy $\pi_{\bm{\theta}}$ is updated only when the buffer reaches a fixed rollout length $T_{rollout}$. This design is aimed to prevent the mixing of semantically heterogeneous trajectories and allow the reinforcement signal to remain consistent with each agent’s trait-driven preferences, facilitating stable and interpretable policy learning across a diverse agent population.

\subsection{Calibration of Agent Trait Distribution}

Since agent traits shape market-level behavior, it is crucial to calibrate their population-level distribution such that the simulated data resembles empirical data. To achieve this, we adopt the OT-based method proposed by \citet{ot_based_simulation_evaluation}, which quantitatively compares the simulation outputs and empirical data. We begin by defining points $\bm{x}_k\in\mathbb{R}^d,~ k\in\{1,\ldots,K\}$, where each $\bm{x}_k$ is a $d$-dimensional feature vector sampled from stock price series data. These points form a point cloud, represented as $X={}^\top\!\begin{pmatrix}\bm{x}_1 & \ldots & \bm{x}_K\end{pmatrix}\in\mathbb{R}^{K\times d}$, where $K$ denotes the number of points. The OT distance between the two point clouds $X_{syn}\in\mathbb{R}^{K\times d}$ and $X_{real}\in\mathbb{R}^{L\times d}$, is defined as:
{\small
\begin{align}
\MoveEqLeft OT(X_{\mathrm{syn}}, X_{\mathrm{real}}) = \min_{P \in \mathbb{R}^{K \times L}}\sum_{k=1}^K \sum_{l=1}^L \left\| X_{\mathrm{real}}^k - X_{\mathrm{syn}}^l \right\|_2^2 P^{k,l}\nonumber \\
\text{s.t.}~~ & \forall k, l~~ 0 \leq P^{k,l}, ~~ \sum_{k=1}^K P^{k,l} = \tfrac{1}{L}, ~~ \sum_{l=1}^L P^{k,l} = \tfrac{1}{K}
\end{align}}
where $X_{syn}$ and $X_{real}$ are the point clouds derived from synthetic and real data, respectively. We denote the $k$-th row element of the point cloud $X$ as $X^k$. $P^{k,l}$ denotes the $k$-th row and $l$-th column element of the matrix $P\in\mathbb{R}_+^{K\times L}$. We define three types of points, $\bm{x}^r, \bm{x}^t\in\mathbb{R}^{1}$, and $\bm{x}^{as}\in\mathbb{R}^9$ to see the underlying order of return distribution, tail return distribution, and absolute return time series in the synthetic data. Let us define a series of one-minute log-return data, arranged in chronological order at equal intervals, and standardized to have a sample mean of $0$ and a sample standard deviation of $1$. Let this standardized series be denoted as $\left((r_{t,i})_{t=1}^{T_{len}}\right)_{i=1}^{N_{days}}$, where $T_{len}$ represents the length of each time series, and $N_{days}$ denotes the total number of collected series. Using $r_{t,i}$, the points are represented as:
{\small
\begin{align}
    \begin{split}
    \MoveEqLeft\bm{x}^r=\begin{pmatrix}r_{t,i}\end{pmatrix},~~~\bm{x}^t=\begin{pmatrix}\tilde{r}_{(k)}\end{pmatrix},\\
    \MoveEqLeft\bm{x}^{as}={}^\top\!\begin{pmatrix}|r_{t,i}| & |r_{t+1,i}| & |r_{t+10,i}| & \ldots & |r_{t+70,i}|\end{pmatrix}
    \end{split}
\end{align}}
where $\tilde{r}_{(k)}$ is calculated as follows. Let $r_{(N)},\ldots,r_{(1)}$ denote the descending order statistics of samples of absolute log-returns $|r_1|,\ldots,|r_N|$ of size $N~(=T_{len}N_{days})$,
{\small
\begin{eqnarray}
\tilde{r}_{(k)}=\log \frac{r_{(N-k+1)}}{r_{(N-K)}},~ k\in\{1,\ldots,K\}
\end{eqnarray}}
$\tilde{r}_{(k)}$ was introduced to see the underlying order of the tail distribution of log-returns, drawing from \citet{hill}. We define the OT distances between real and synthetic point clouds computed using the points $\bm{x}^r$, $\bm{x}^t$, and $\bm{x}^{as}$ as $OT^r$, $OT^t$, and $OT^{as}$, respectively. During calibration, the agents with each candidate combination of $\lambda^{\sigma}$, $\lambda^{\alpha}$, and $\lambda^{\gamma}$ are independently trained, and compared with weighted average of OT distances $\bar{OT}=\omega^rOT^r+\omega^tOT^t+\omega^{as}OT^{as}$ 
where $\omega^r,\omega^t,\omega^{as}\in\mathbb{R}_+$ are the weights assigned to each distance.

\section{Experiment}
Through the experiment, we conduct market simulations using agents trained by our method, and investigate the following two research questions: \textbf{RQ1}: How does agent heterogeneity in preferences lead to behavioral differentiation? and \textbf{RQ2}: How does it lead to collective dynamics? For RQ1, we analyze the relationship between the incorporated agent traits and their behavioral patterns. For RQ2, we evaluate the realism of the simulated outcomes by comparing them against baselines where either learning or heterogeneity in preferences is modeled. Finally, we  discuss the emergent mechanisms of stock market dynamics from these analyses.

In the experiment, we searched the hyperparameters $\lambda^\sigma$, $\lambda^\alpha$, and $\lambda^\gamma$ using OT-based calibration method. For real data, we used FLEX-FULL historical tick data~\citep{flex_full}.

\subsection{Baselines}
We compare the realism of the simulation results between our method and the following baseline ABMs.
\begin{itemize}
\item \textbf{Zero intelligence (ZI-) Agent}~\citep{zero_intelligence}: In financial market simulations, ZI-Agents serve as a behaviorally naive benchmark. They submit one unit buy or sell orders randomly, without any optimization or strategic reasoning.
\item \textbf{Fundamental-chartist-noise (FCN-) Agent}~\citep{fcn2}: FCN-Agent is a widely adopted agent model for LOB market simulations. It is primarily a mixture of fundamental and chartist investment strategies, incorporating heterogeneity in time horizons and risk aversion.
\item \textbf{adaptive FCN-Agent (adFCN-Agent)}~\citep{adaptive_fcn_agent}: adFCN-Agent is a variant of FCN-Agent, which adaptively selects between fundamental and chartist strategies based on their recent predictive accuracy.
\item \textbf{Ours (fixed)}: Our method without trait heterogeneity. All agents' trait factors are fixed with population mean.%as: $\forall j~~\alpha^j=2.0,\sigma^j=0.02,\gamma^j=0.95$.
\end{itemize}
The FCN-Agent models only heterogeneity in preferences, whereas the adFCN-Agent with fixed risk aversion and time horizons across agents (adFCN-Agent (fixed)) models only learning. The key parameters of these ABMs were calibrated using the OT distance $\bar{OT}$, in the same manner as in our model.

\begin{figure}[bp]
  \centering
  \includegraphics[width=0.8\linewidth]{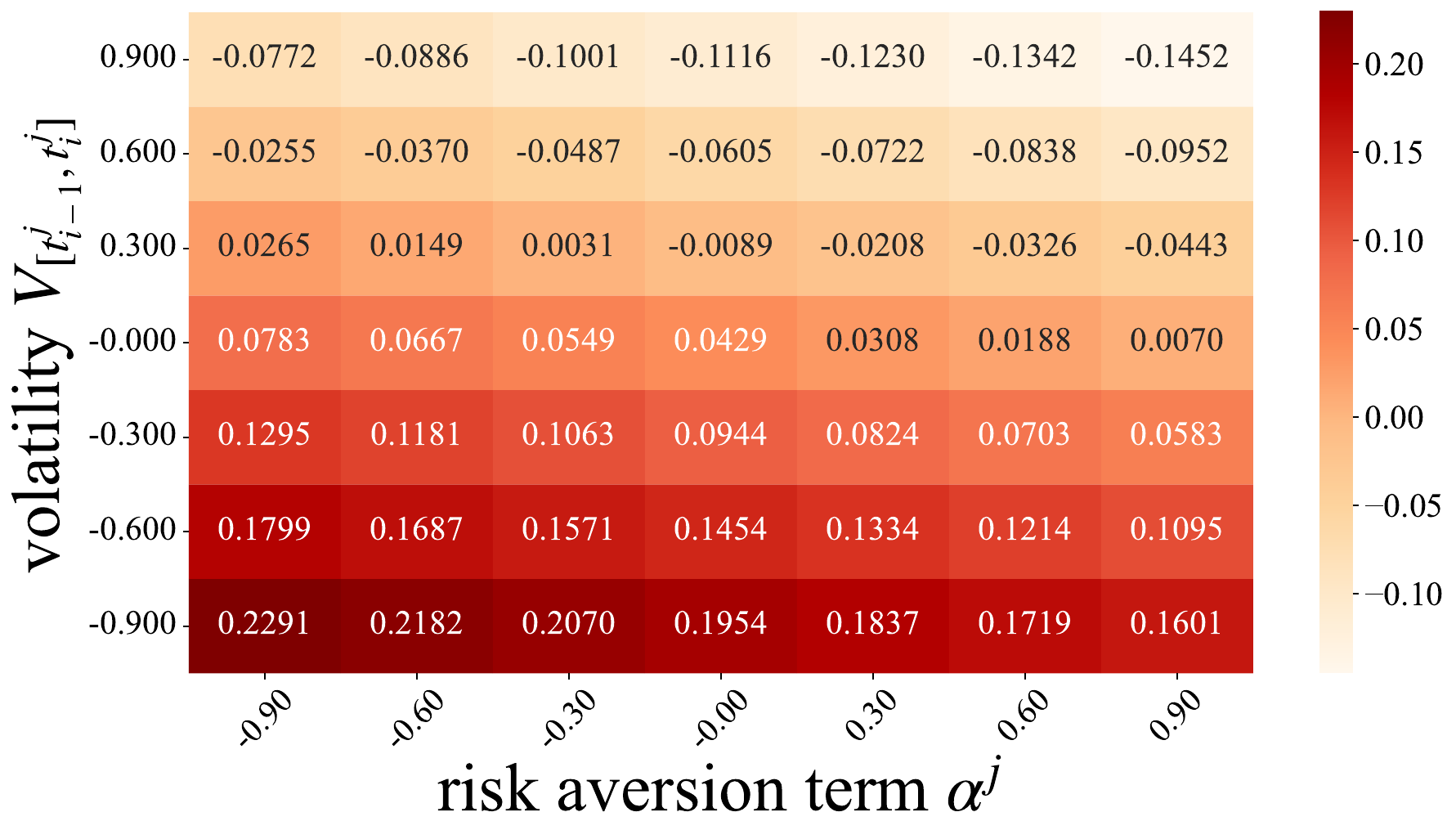}
  \caption{Heatmap of the scaled order volume $\tilde{v}_t^j$ derived from the obtained policy, with varying rescaled volatility $V_{[t_{i-1}^j, t_i^j]}$ and risk aversion term $\alpha^j$.}
  \label{Fig:heatmap_order_volume_given_alpha_volatility}
\end{figure}

\subsection{Evaluation Metrics}
In addition to the OT distances, we assessed whether the synthetic data exhibited key stylized facts observed in real financial markets. Following \citet{stylized_facts} and \citet{realism}, we considered the following stylized facts:

\begin{itemize}
\item \textbf{Kurtosis}: The kurtosis of the stock return distribution is positive, indicating fat tails.
\item \textbf{Tail coefficient (Tail coef.)}: The Hill tail exponent~\cite{hill} is approximately three, reflecting the power-law behavior of extreme returns.
\item \textbf{Autocorrelation coefficient (Acorr coef.)}: The estimated coefficient $\hat{\zeta}_r^{\mathrm{acorr}}$ in the linear regression of $\log \mathrm{Corr}(|r_{t,i}|, |r_{t+\tau,i}|)$ on $\log \tau$ lies between 0 and 1, indicating the long memory property of absolute return series. Here, $\mathrm{Corr}(\cdot,\cdot)$ denotes the correlation operator.
\item \textbf{Volume-Volatility correlation (VV corr.)}: The correlation between the executed trading volume %within a one-minute interval
and the absolute return is positive. %over the same interval is positive.
\end{itemize}
\begin{figure}[tbp]
  \centering
  \includegraphics[width=\linewidth]{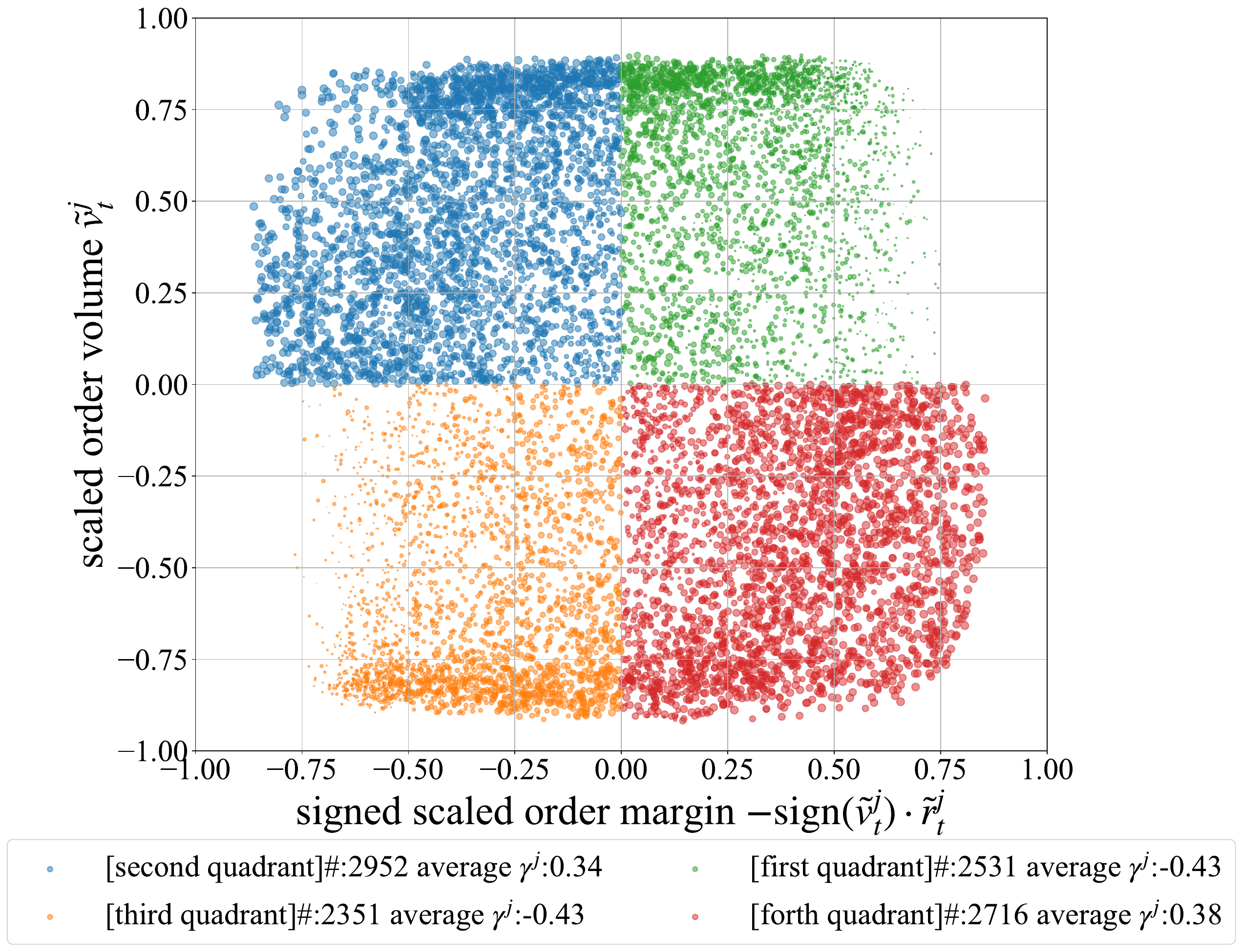}
  \caption{Scatter plot of agent action vectors from five simulations, with color coding for each quadrant. The size of each point represents the agent's discount factor $\gamma^j$.}
  \label{Fig:scatter_actions}
\end{figure}

\section{Results and Discussions}

This section presents and discusses our experimental results. RQ1 analyzes the micro-level agent strategies and the meso-level ecological niches they form, while RQ2 examines the macro-level structures that drive collective market dynamics.

\subsection{RQ1: Behavioral Differentiations}% Driven by Heterogeneous Preference}

\subsubsection{Behavioral Analyses}

To investigate RQ1, we analyzed the learned policy obtained through our method to investigate how each trait factor contributes to shaping agent behavioral patterns. Figure~\ref{Fig:heatmap_order_volume_given_alpha_volatility} presents a heatmap illustrating the values of $\tilde{v}_t^j$ while varying the two observation components, $V_{[t_{i-1}^j, t_i^j]}$ and $\alpha^j$. The values of the remaining variables were fixed. Figure~\ref{Fig:heatmap_order_volume_given_alpha_volatility} indicates that agents tended to sell stocks more actively as their risk aversion increased and market volatility rose.

Figure~\ref{Fig:scatter_actions} presents the scatter plot of agent actions $\tilde{v}_t^j$ and $\tilde{r}_t^j$ across five simulation trials. Agent actions are distributed across all quadrants of the action space, defined by order volume and margin. The second and fourth quadrants represent behavior consistent with price-sensitive trading---selling above the mid price or buying below it---while the first and third quadrants reflect the opposite: trading at disadvantageous prices. The average discount factors $\gamma^j$ differ systematically across these regions. Agents acting in the second and fourth quadrants tend to have higher $\gamma^j$ values, suggesting a more patient, forward-looking attitude. Conversely, agents in the first and third quadrants exhibit lower average discount factors, indicating a more myopic preference structure. This pattern suggests that myopic agents are more willing to accept unfavorable prices to execute their trades immediately, prioritizing execution speed over price optimization. These findings indicate that the learned shared-policy is not mere averaged compromises, but rather reflect meaningful behavioral differentiation driven by agent traits.

\begin{figure}[tbp]
  \centering
  \includegraphics[width=0.9\linewidth]{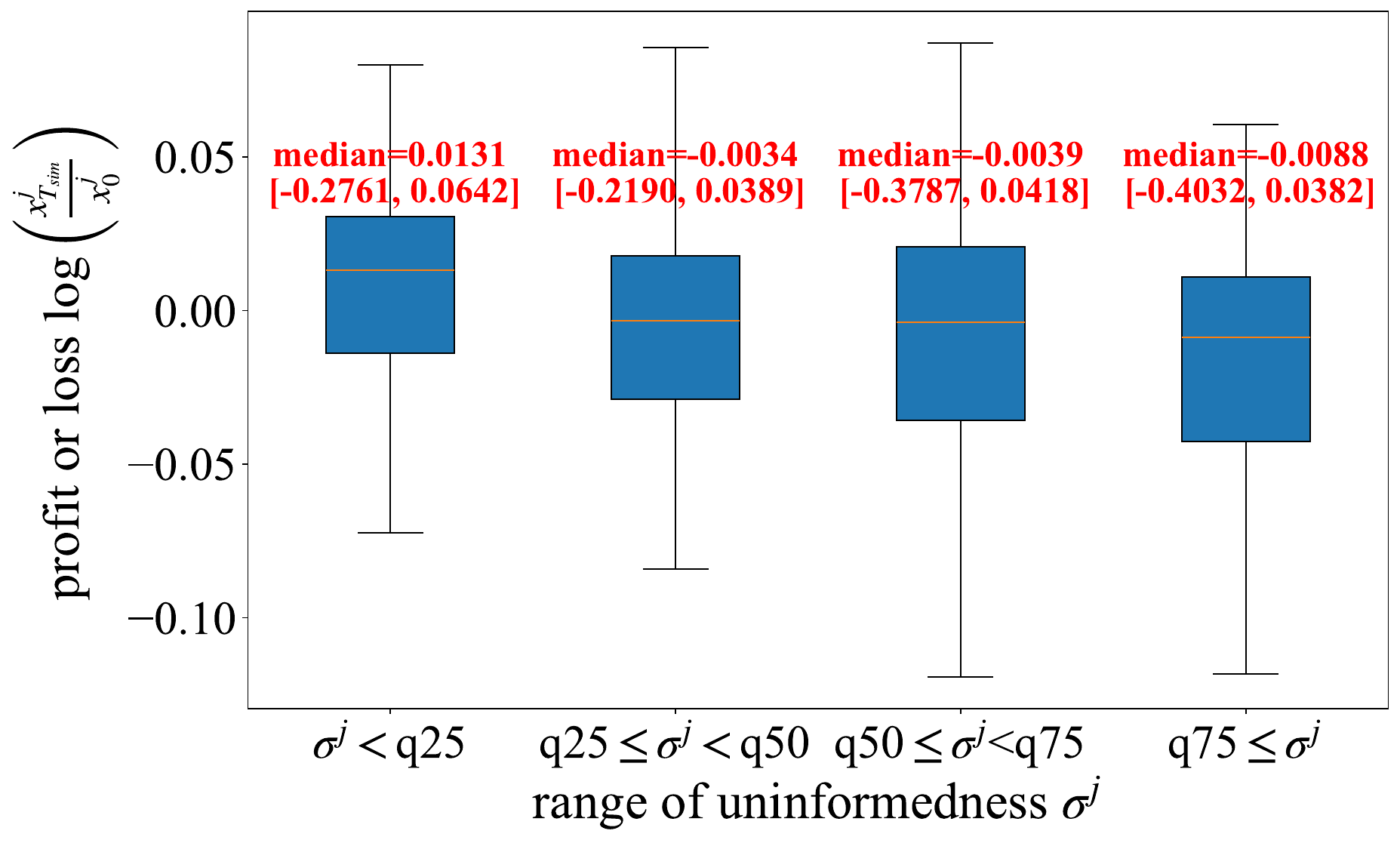}
  \caption{Box plot of agents’ trading performance over five runs. Agents are grouped by uninformedness $\sigma^j$ quartiles (q{quartile(\%)}) on the horizontal axis, and vertical axis shows log returns. %$\log x_{T_{sim}}^j - \log x_0^j$.
  Boxes indicate medians and 90\% ranges, showing the distribution of each trading performance.}
  \label{Fig:boxplot_uninformedness}
\end{figure}

Figure~\ref{Fig:boxplot_uninformedness} presents a box plot illustrating the relationship between agent uninformedness $\sigma^j$ and trading performance. Agents are grouped into quartiles based on their level of uninformedness, with lower $\sigma^j$ indicating better access to information. The plot reveals a negative trend: as uninformedness increases, the profit distribution shifts downward. This pattern demonstrates how differences in access to information influence agent behavior and ultimately lead to disparities in trading performance.

\begin{figure*}[t]
  \centering
  % ---- 1st row (4 figs) ----
  \begin{subfigure}[b]{0.24\textwidth}
    \includegraphics[width=\linewidth]{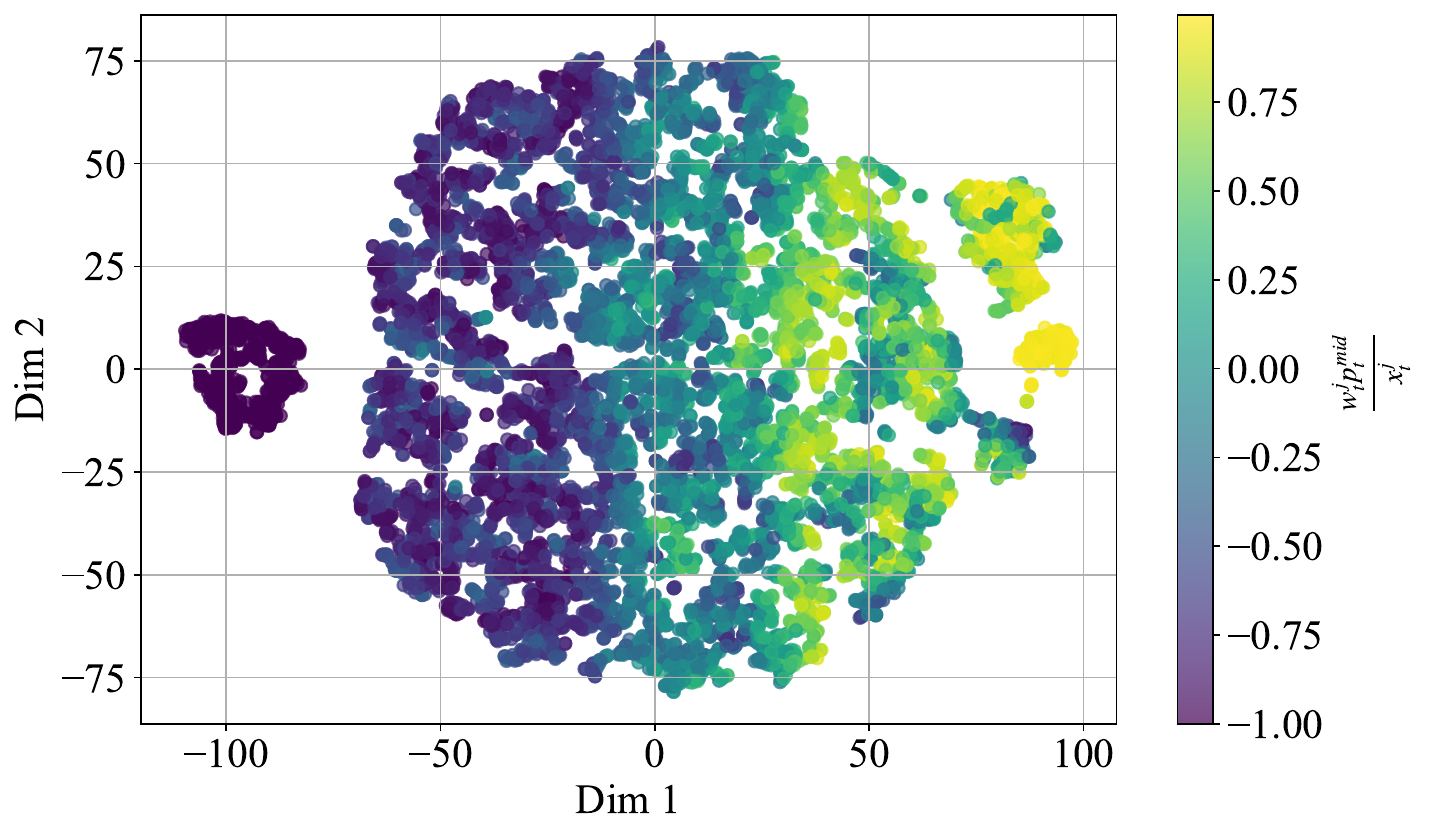}
    \caption{Holding asset ratio: $\frac{w_t^j p_t^{mid}}{x_t^j}$}
  \end{subfigure}
  \begin{subfigure}[b]{0.24\textwidth}
    \includegraphics[width=\linewidth]{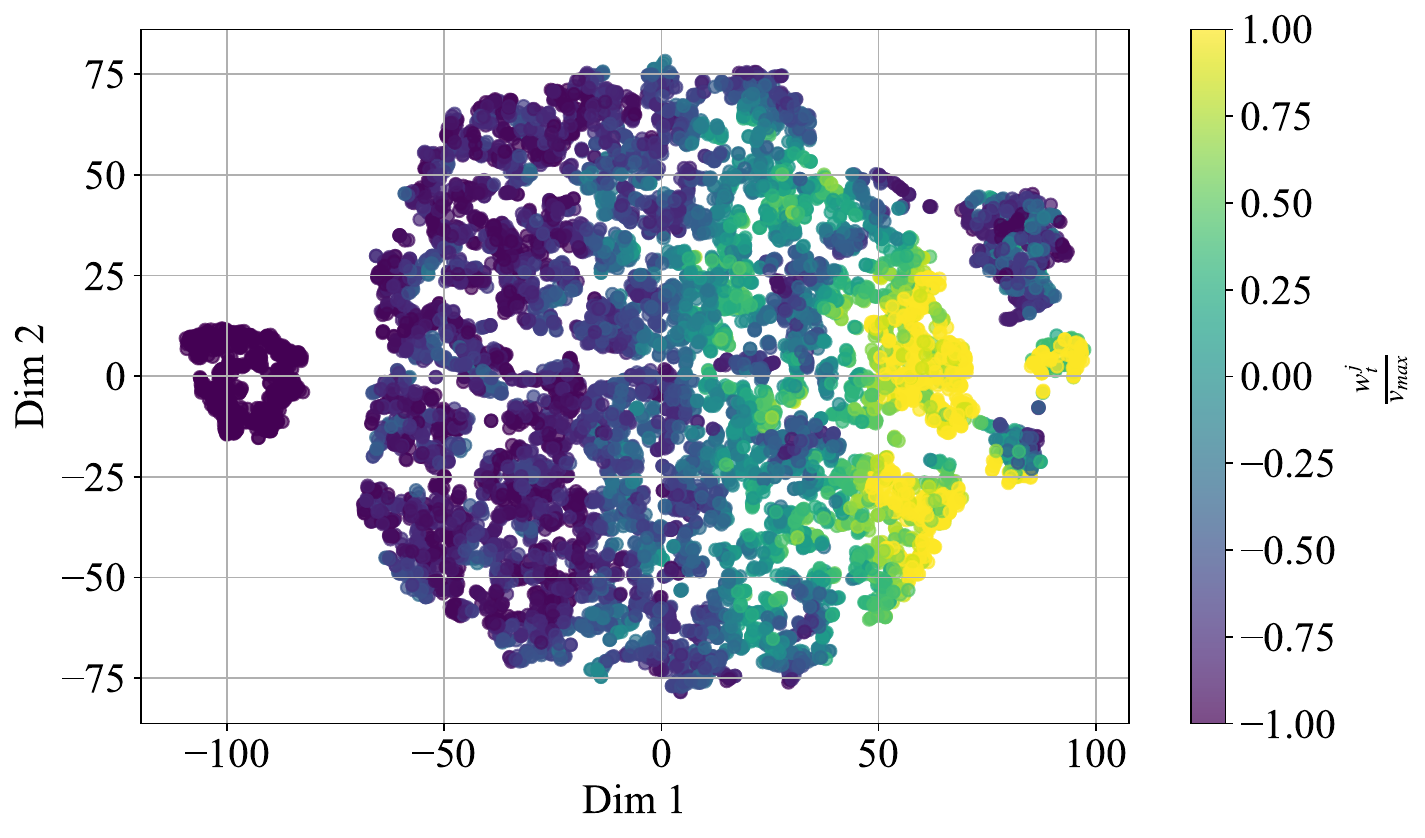}
    \caption{Holding asset-to-maximum order volume ratio: $\frac{w_t^j}{v_{max}}$}
  \end{subfigure}
  \begin{subfigure}[b]{0.24\textwidth}
    \includegraphics[width=\linewidth]{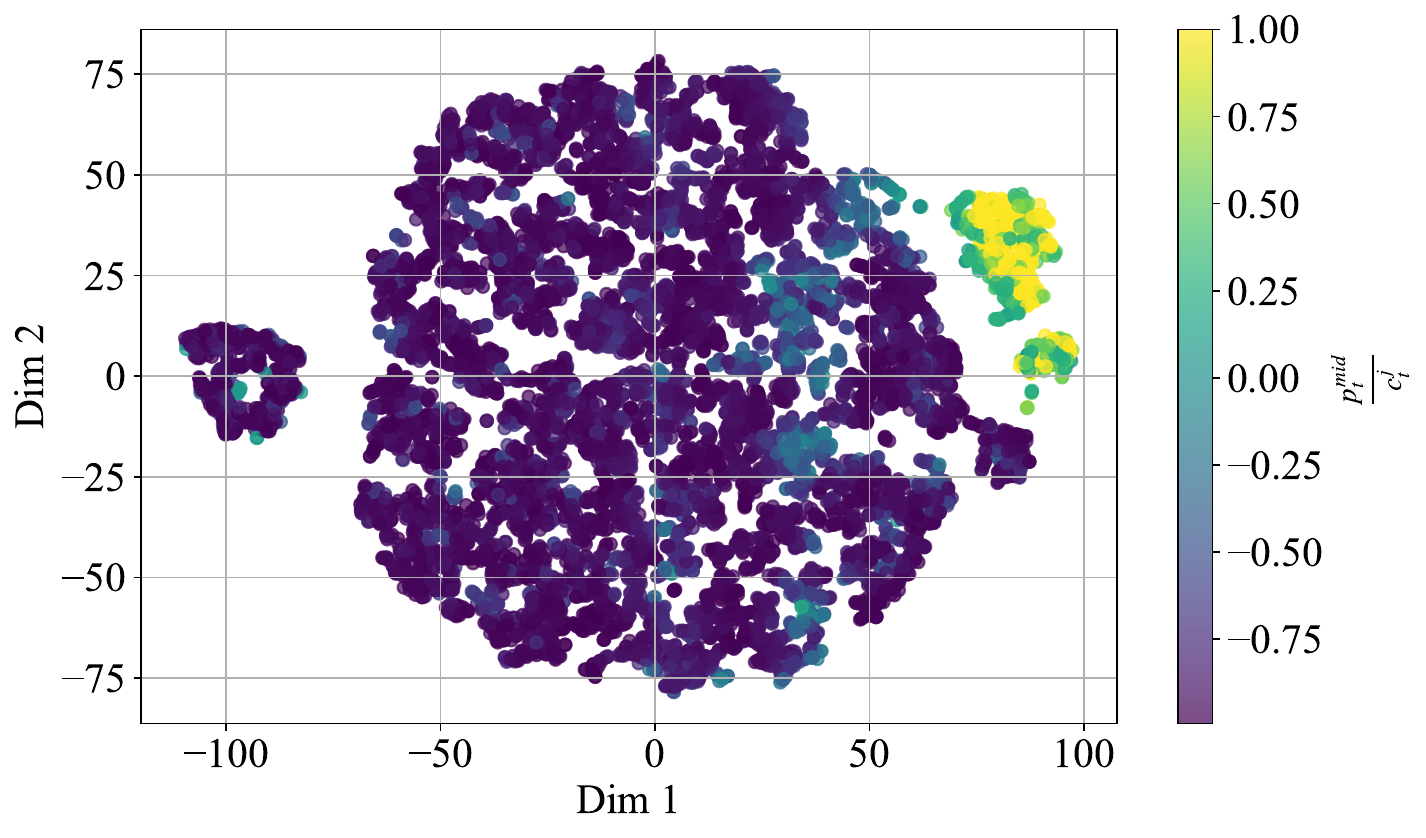}
    \caption{Inverted buying power: $\frac{p_t^{mid}}{c_t^j}$}\label{Fig:tsne_inverted_buying_power}
  \end{subfigure}
  \begin{subfigure}[b]{0.24\textwidth}
    \includegraphics[width=\linewidth]{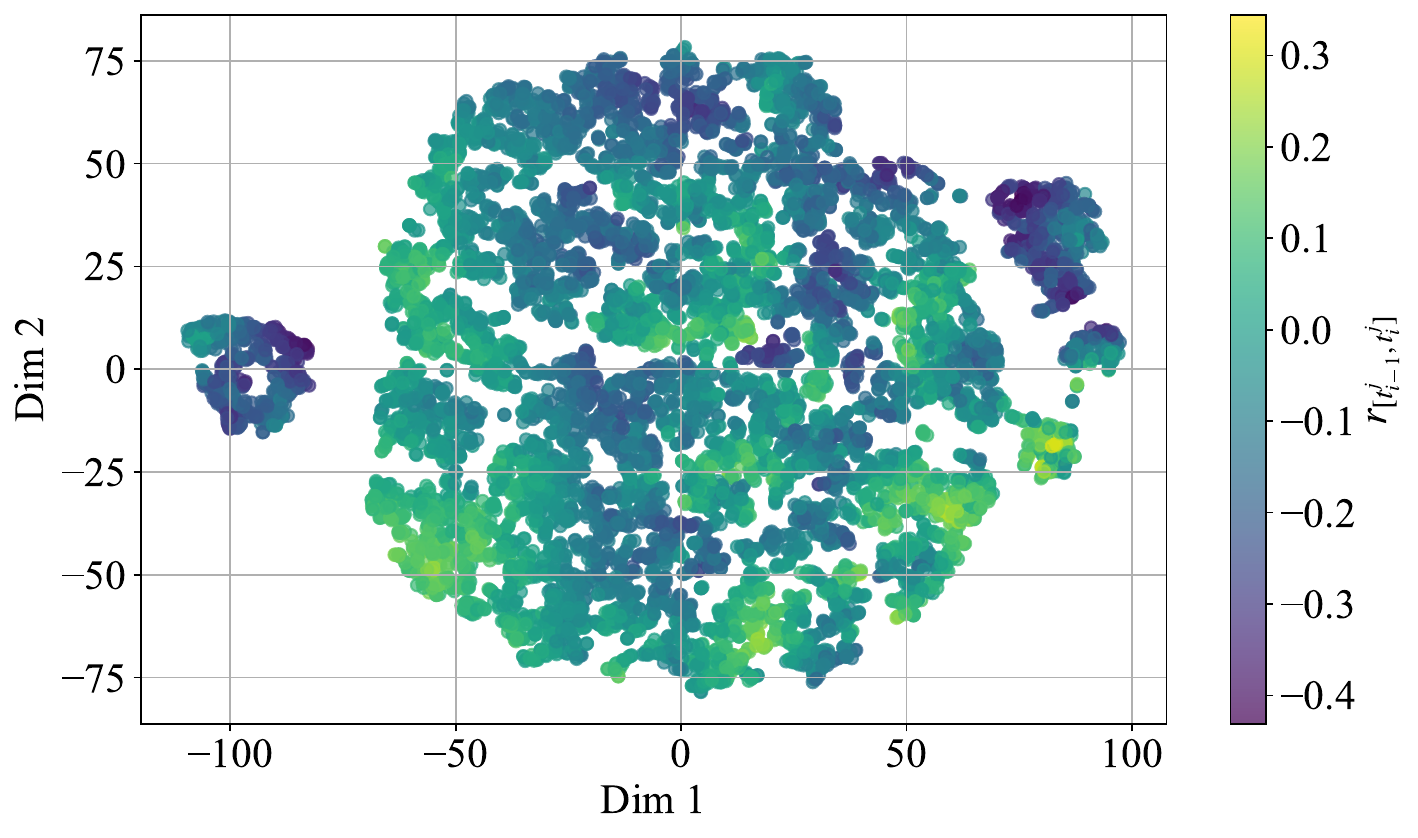}
    \caption{Return: $r_{[t_{i-1}^j, t_i^j]}$}
  \end{subfigure}

  % ---- 2nd row (4 figs) ----
  \begin{subfigure}[b]{0.24\textwidth}
    \includegraphics[width=\linewidth]{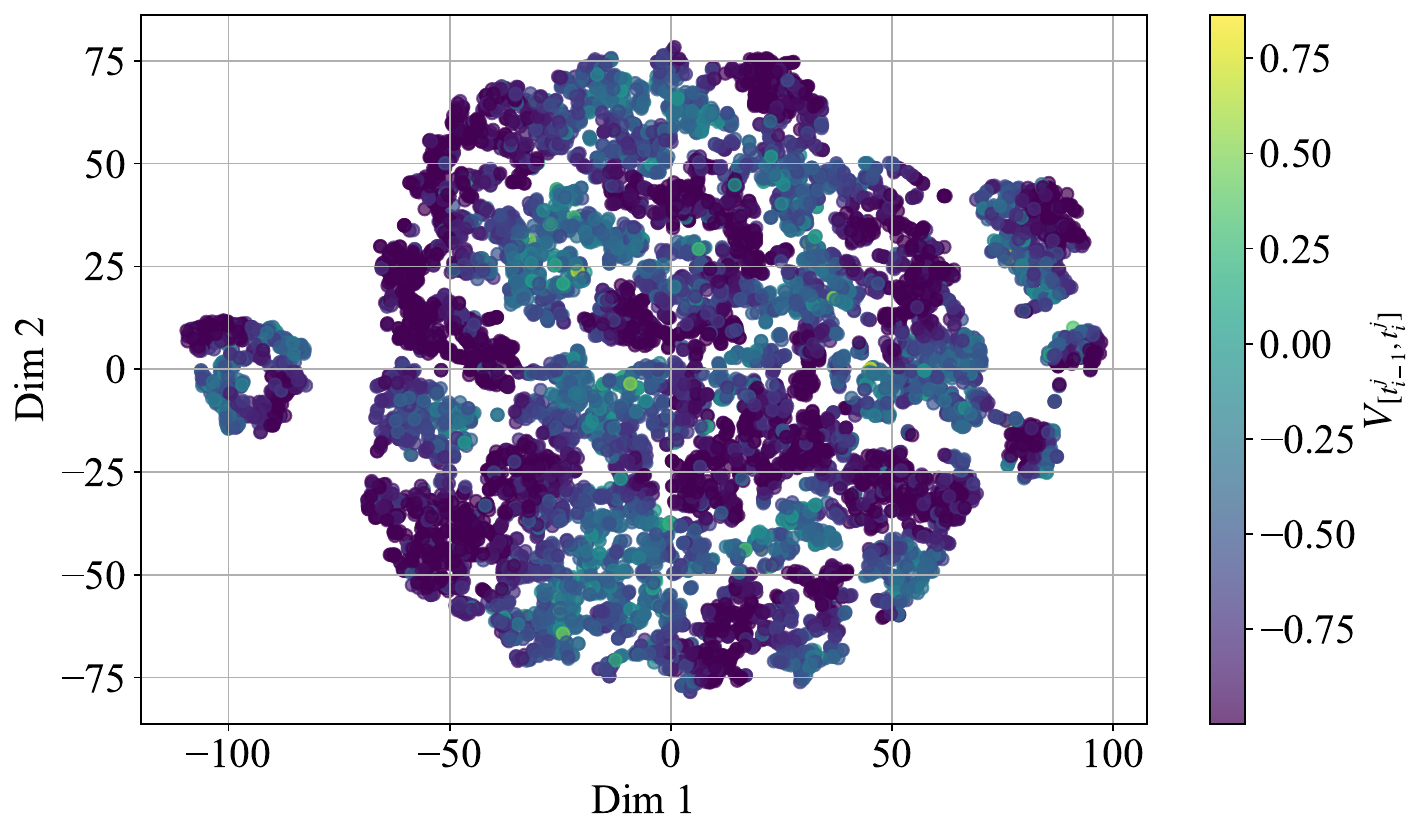}
    \caption{Volatility: $V_{[t_{i-1}^j, t_i^j]}$}
  \end{subfigure}
  \begin{subfigure}[b]{0.24\textwidth}
    \includegraphics[width=\linewidth]{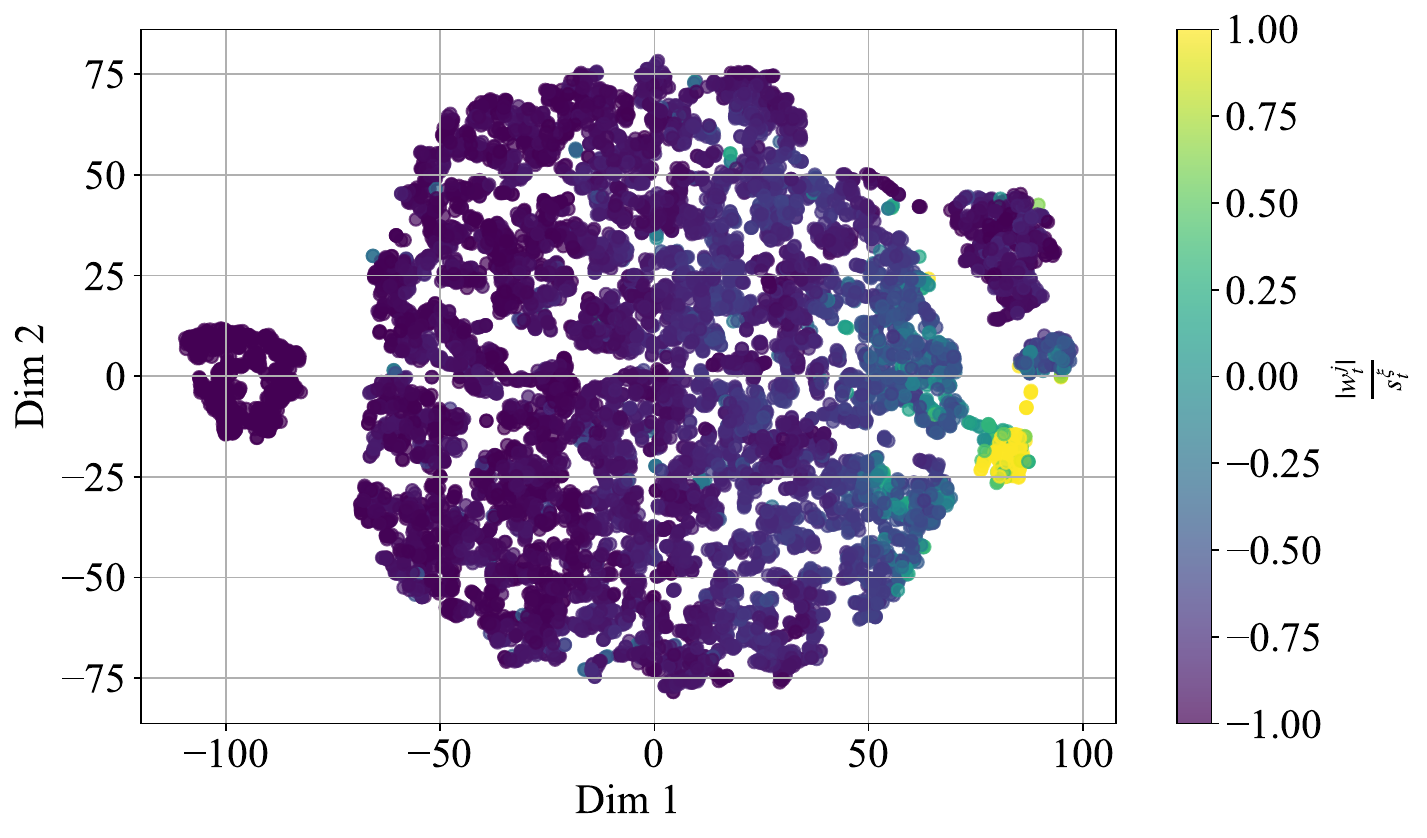}
    \caption{Asset volume-to-buy order ratio: $\frac{|w_t^j|}{b_t^\xi}$}\label{Fig:tsne_asset_volume_buy_orders}
  \end{subfigure}
  \begin{subfigure}[b]{0.24\textwidth}
    \includegraphics[width=\linewidth]{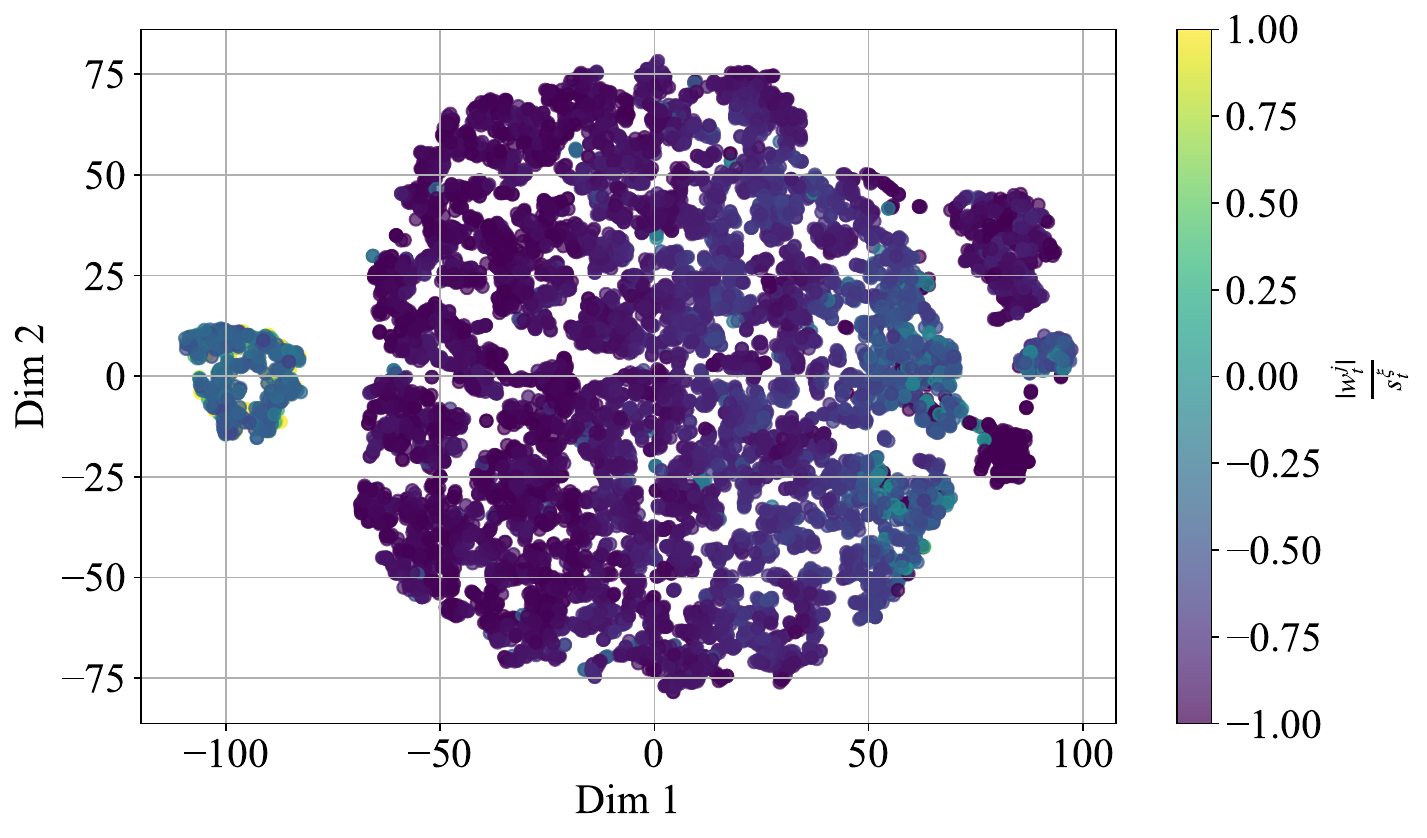}
    \caption{Asset volume-to-sell order ratio: $\frac{|w_t^j|}{s_t^\xi}$}\label{Fig:tsne_asset_volume_sell_orders}
  \end{subfigure}
  \begin{subfigure}[b]{0.24\textwidth}
    \includegraphics[width=\linewidth]{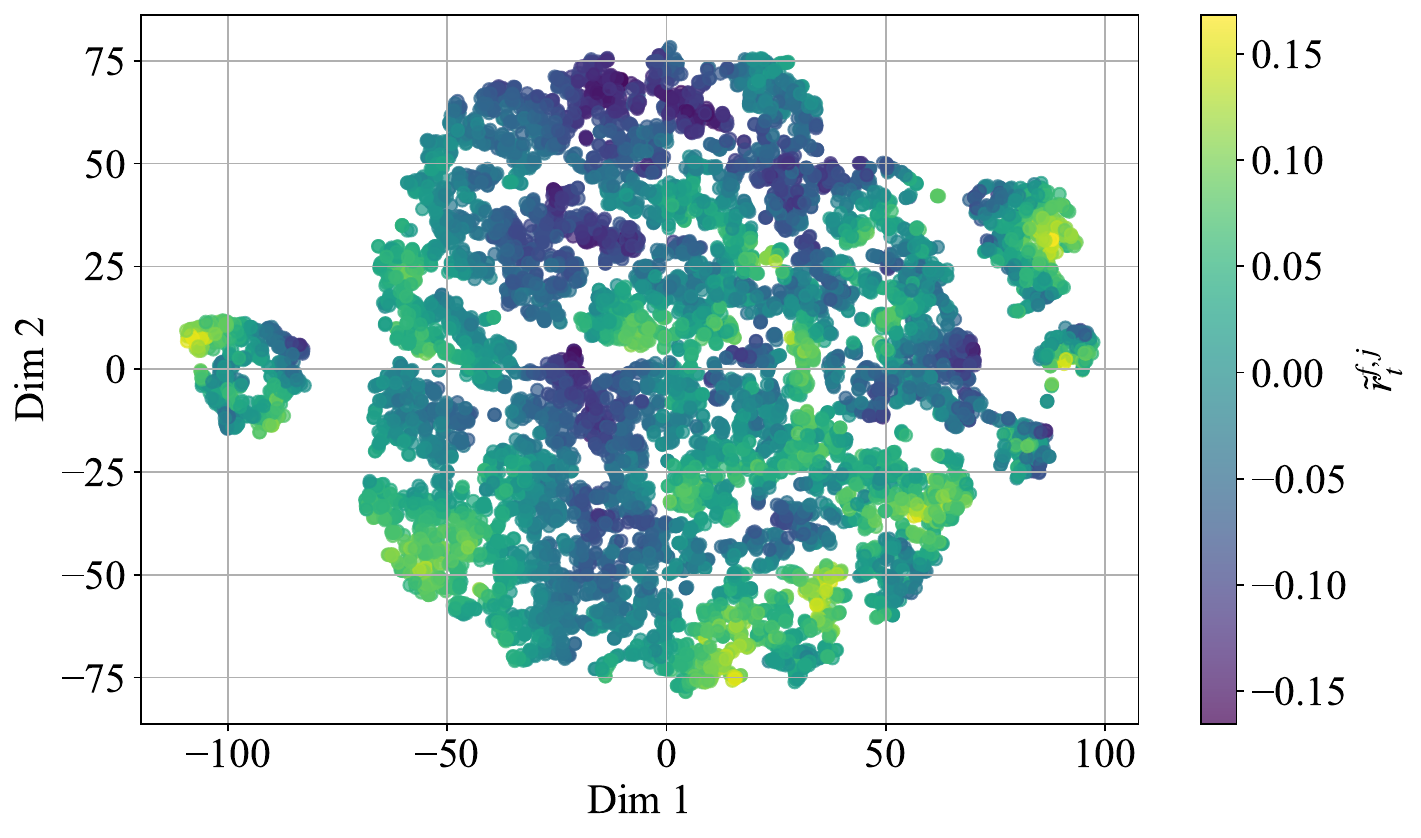}
    \caption{Blurred fundamental return: $\tilde{r}_t^{f,j}$}
  \end{subfigure}

  % ---- 3rd row (3 figs) ----
  \begin{subfigure}[b]{0.24\textwidth}
    \includegraphics[width=\linewidth]{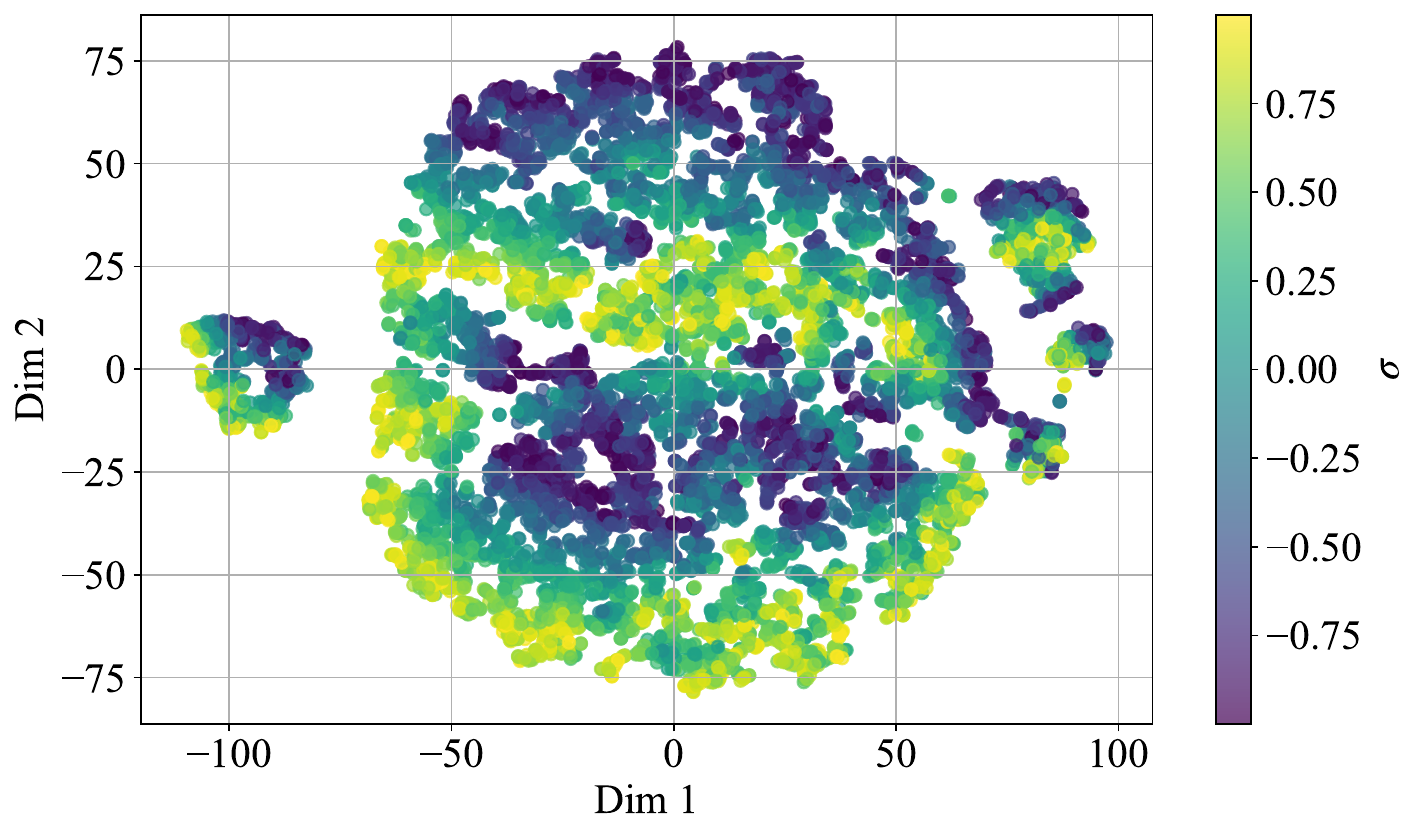}
    \caption{Uninformedness: $\sigma^j$}\label{Fig:tsne_sigma}
  \end{subfigure}
  \begin{subfigure}[b]{0.24\textwidth}
    \includegraphics[width=\linewidth]{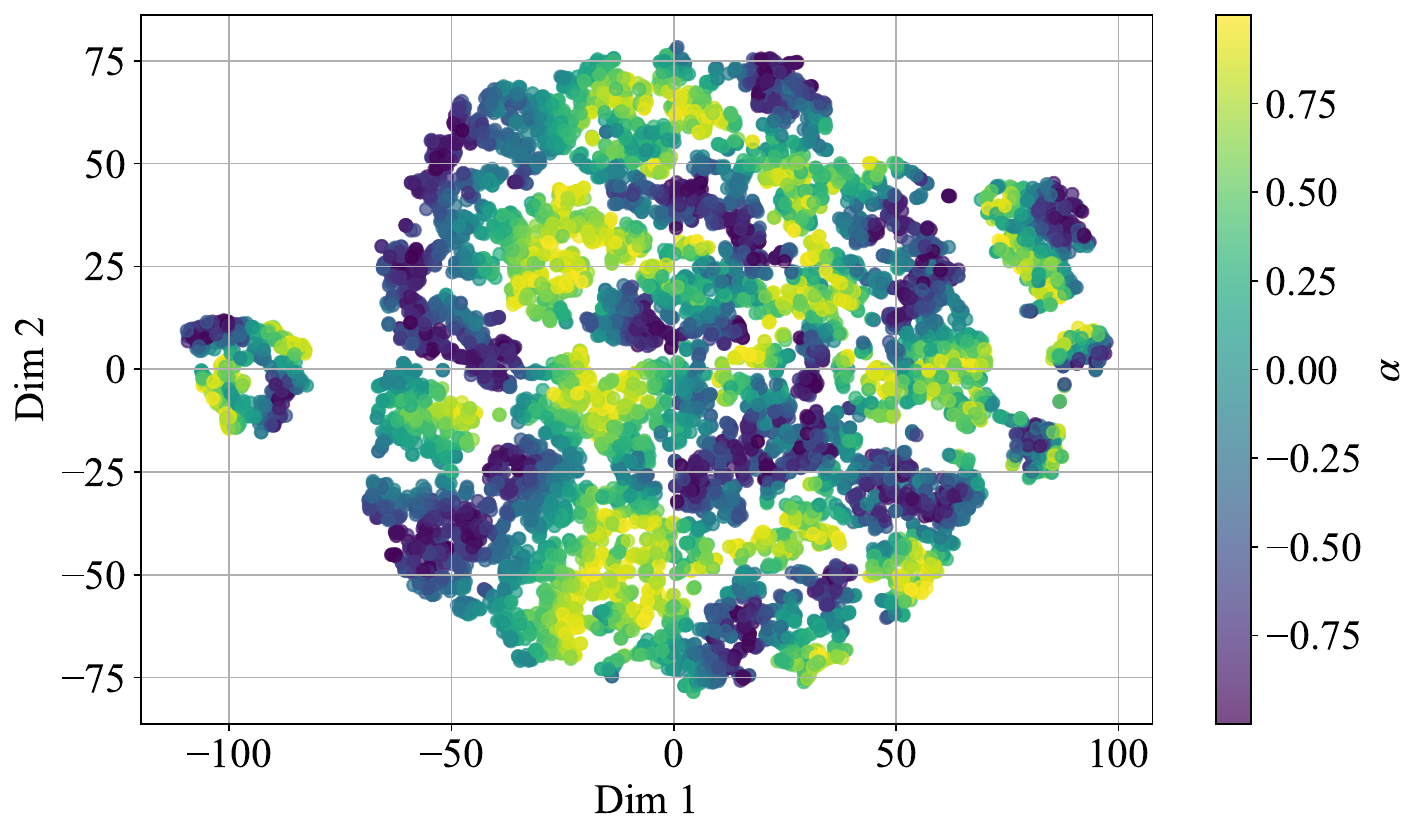}
    \caption{Risk aversion: $\alpha^j$}\label{Fig:tsne_alpha}
  \end{subfigure}
  \begin{subfigure}[b]{0.24\textwidth}
    \includegraphics[width=\linewidth]{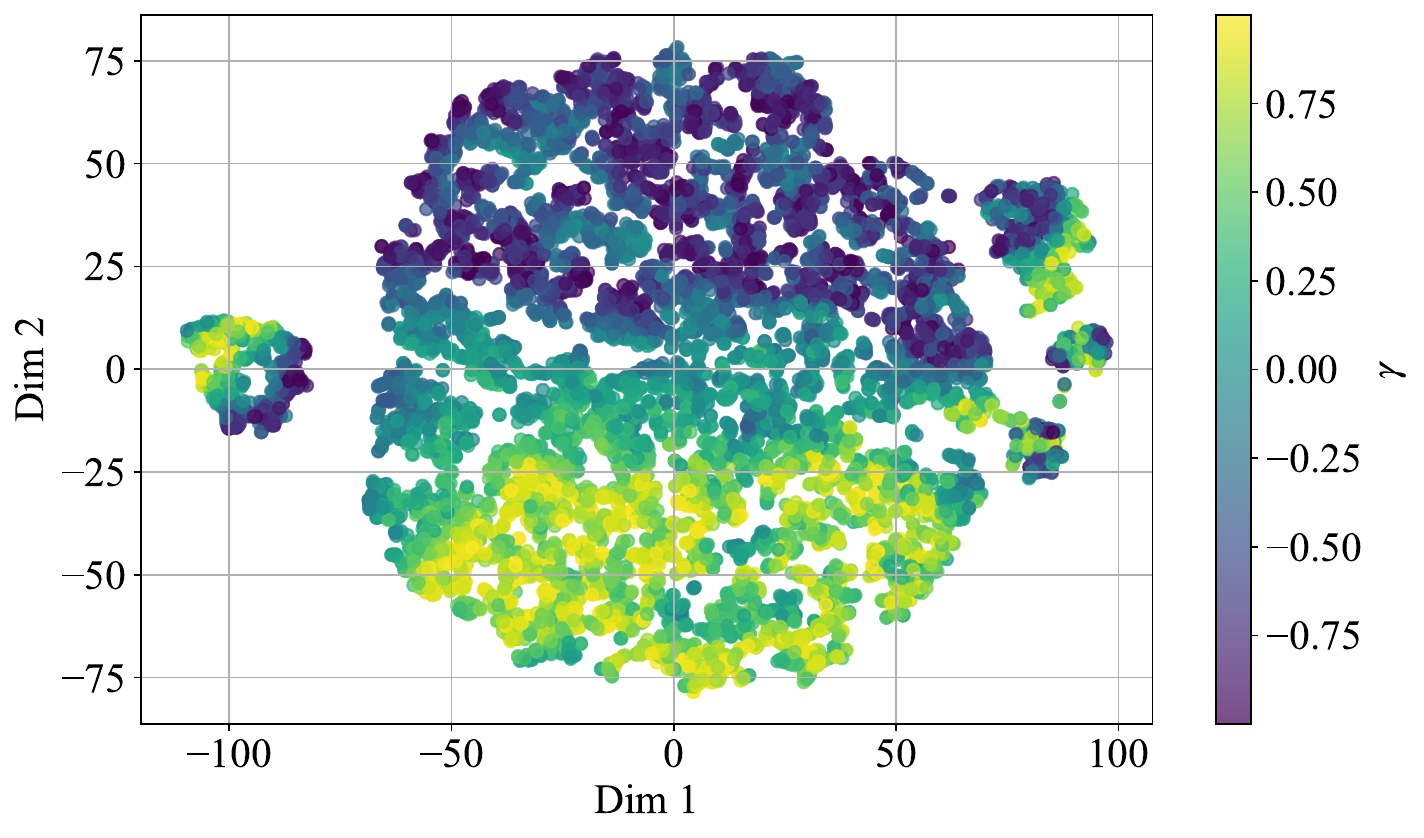}
    \caption{Discount factor: $\gamma^j$}\label{Fig:tsne_gamma}
  \end{subfigure}

  \caption{t-SNE visualizations of internal representations in the second hidden layer, colored by various observation components.}
  \label{Fig:tsne_observation_components}
\end{figure*}

\subsubsection{Probing of the Shared-Policy}

To further support our claim to answer RQ1 that agent behaviors are differentiated according to their individual traits, we conducted a probing analysis on the learned shared-policy network. Figures~\ref{Fig:tsne_observation_components} show the t-SNE visualizations of the internal representations extracted from the second hidden layer of the shared-policy network. These representations were collected across five independent evaluation episodes. These visualizations illustrate how different observation components are embedded in the learned internal representations. As shown in Figures~\ref{Fig:tsne_observation_components}~(\subref{Fig:tsne_sigma}, \subref{Fig:tsne_alpha}, \subref{Fig:tsne_gamma}), the internal representations are systematically organized according to agent-specific traits. In the three visualizations, the data points are distinctly clustered by their corresponding trait values. This consistent trait-wise separation in the second hidden layer suggests that the network implicitly encodes a latent notion of {\em who the agent is} based on its individual preferences.

Figures~\ref{Fig:tsne_observation_components}~(\subref{Fig:tsne_sigma}, \subref{Fig:tsne_alpha}, \subref{Fig:tsne_gamma}) show smooth, continuous gradients along the latent dimensions when colored by each agent’s trait parameters, indicating that the shared-policy develops an internal representation that smoothly encodes who the agent is---i.e., its identity defined by long-lasting traits. In contrast, Figures~\ref{Fig:tsne_observation_components}~(\subref{Fig:tsne_inverted_buying_power}, \subref{Fig:tsne_asset_volume_buy_orders}, \subref{Fig:tsne_asset_volume_sell_orders}) reveal more clustered and discrete structures. These features correspond to the agent’s transient state, such as its current wealth or market liquidity. Notably, the reward function includes components like the cash shortage penalty and illiquidity penalty, which are implemented as piecewise constant functions---yielding sharply negative rewards when certain thresholds are crossed. This structural design likely induces discrete transitions in the agent’s internal representation of situational context. Together, these results suggest that the learned shared-policy not only encodes each agent’s stable identity (trait-driven preferences) in a smooth manner, but also dynamically reacts to short-term agent or market states in a thresholded, categorical fashion. In doing so, the policy integrates both long-term traits and short-term states to adaptively adjust agent behavior, supporting the emergence of diverse and context-aware strategies.

\begin{table}[tbp]
\centering
\caption{Ablation study on preference heterogeneity. We report the average of the discounted cumulative utility $\mathcal{U}$ (mean $\pm$ std. over five trials). }
\begin{tabular}{cc}
\toprule
Model & Avg. Cum. Disc. Utility\\
\midrule
Ours & $1016.88~(\pm41.22)$ \\
Ours (homo-$\alpha^j$) & $846.58~(\pm29.85)$ \\
Ours ($\alpha^j$ masked) & $799.32~(\pm83.47)$ \\
Ours (homo-$\gamma^j$) & $861.98~(\pm51.02)$ \\
Ours ($\gamma^j$ masked) & $606.58~(\pm102.39)$ \\
Ours (homo-$\alpha^j$) & $1009.98~(\pm44.00)$ \\
Ours ($\alpha^j$ masked) & $916.53~(\pm20.87)$ \\
\bottomrule
\end{tabular}
\label{Tab:ablation_heterogeneity}
\end{table}

\subsubsection{Ablation Study}

\begin{table*}[tb]
\centering
\caption{Comparison of key stylized facts and OT distances from real data across financial ABMs. A check mark ($\checkmark$) denotes conformity to each stylized fact. OT columns show mean distances between synthetic and real point clouds (standard deviations in parentheses); the best values are in \textbf{bold}.}
\begin{tabular*}{\textwidth}{@{\extracolsep{\fill}}lccccccc}
\toprule
          & \multicolumn{4}{c}{Stylized Facts} & \multicolumn{3}{c}{OT Distances from Real Data} \\ 
\cmidrule(lr){2-5} \cmidrule(lr){6-8}
          & Kurtosis & Tail coef. & Acorr coef. & VV corr. & $OT^r~(\times10^{-1})$ & $OT^t~(\times10^{-2})$ & $OT^{as}$ \\
\midrule
 Real
 data     & 7.79$\checkmark$ & 2.96$\checkmark$ & 0.71$\checkmark$ & 0.43$\checkmark$ & \textemdash{} & \textemdash{} & \textemdash{} \\
\midrule
ZI-Agent & 1.19$\checkmark$ & 3.79 & \textemdash{} & 0.04$\checkmark$ & $1.80~(\pm0.79)$ & $1.00~(\pm0.52)$ & $0.34~(\pm0.03)$\\
FCN-Agent & 6.25$\checkmark$ & 3.23 & 1.11 & 0.08$\checkmark$ & $1.34~(\pm0.31)$ & $0.91~(\pm0.61)$ & $0.30~(\pm0.02)$ \\
adFCN-Agent (fixed) & 31.38$\checkmark$ & 2.56 & 1.01 & 0.01$\checkmark$ & $2.34~(\pm0.45)$ & $1.05~(\pm0.62)$ & $0.26~(\pm0.02)$ \\
adFCN-Agent & 10.38$\checkmark$ & 2.82$\checkmark$ & 0.77$\checkmark$ & 0.05$\checkmark$ & $0.65~(\pm0.27)$ & $0.48~(\pm0.30)$ & $0.26~(\pm0.04)$ \\
\midrule
 Ours      & 8.65$\checkmark$ & 2.99$\checkmark$ & 0.92$\checkmark$ & 0.16$\checkmark$ & $\textbf{0.37}~(\pm0.32)$ & $\textbf{0.32}~(\pm0.24)$ & $\textbf{0.10}~(\pm0.03)$ \\
Ours (fixed) & 12.89$\checkmark$ & 3.44 & 1.05 & 0.16$\checkmark$ & $1.40~(\pm0.89)$ & $0.65~(\pm0.56)$ & $0.19~(\pm0.02)$ \\
\bottomrule
\end{tabular*}
\label{Tab:benchmark_results}
\end{table*}

We conducted an ablation study to evaluate the role of preference heterogeneity in sustaining social welfare. Specifically, we ran five independent simulations for each setting and computed the sum of discounted cumulative utilities across all agents, denoted as $\mathcal{U}$:
\begin{align}
\mathcal{U}=\sum_{j=1}^n\sum_{i=1}^{\iota_j}(\gamma^j)^iu_{t_i^j}^j
\end{align}
Two types of ablations were implemented: (i) homo, where the focal trait of agents is fixed to its population mean, thereby eliminating heterogeneity, and (ii) masked, where the corresponding trait input to the shared-policy network is masked.
As reported in Table~\ref{Tab:ablation_heterogeneity}, removing heterogeneity in key preference parameters such as risk aversion ($\alpha^j$) and time discounting ($\gamma^j$) leads to a substantial reduction in overall social utility. This result highlights that when all agents share the same risk attitude or time horizon, the system loses the diversity of demand–supply matching that is otherwise enabled by heterogeneous preferences. For example, risk-tolerant agents and risk-averse agents, or short-term oriented and long-term oriented agents, complement each other in ways that increase aggregate welfare.
The fact that utility decreases in both homo and masked settings provides evidence that preference-driven behavioral differentiation plays a critical role in enhancing welfare. Each agent adapts its strategy in alignment with its individual preference, leading to a diversified financial ecosystem in which agents specialize in distinct niches. Preference heterogeneity, when coupled with learning, induces functional complementarity~\citep{complementarity} among such agent niches---an organismic order in which no single subpopulation can reproduce the system’s performance alone.

\subsection{RQ2: Emergence of Market Dynamics}
To address RQ2, Table~\ref{Tab:benchmark_results} compares our method with baseline ABMs.
FCN-Agent incorporates only heterogeneous preferences, while adFCN-Agent (fixed) and Ours (fixed) model only the learning mechanism. In contrast, adFCN-Agent and Ours implement both heterogeneity and learning simultaneously. The results show that our model and adFCN-Agent successfully reproduce all the key stylized facts, whereas models that adopt only one of the two mechanisms fail to capture certain empirical regularities. This finding suggests that both heterogeneous preferences and adaptive learning are indispensable drivers of collective dynamics in financial markets.
Moreover, the OT distances reveal a clear advantage of our method. Across all three metrics, Ours achieves the lowest distance from real data, outperforming not only the purely heterogeneity-based and purely learning-based baselines, but also the adFCN-Agent. To summarize, the behavioral differentiation aligned with heterogeneous preferences gives rise to realistic dynamics in financial markets through agent interactions. We refer to this hierarchical structure---where micro-level differentiation induces macro-level order---as {\em emergence from emergence}. The superior performance of Ours over adFCN-Agent can be attributed to the description of emergence from emergence, which more accurately reflects how real markets evolve through successive layers of adaptation and organization spanning individual and collective scales.

\begin{figure}[btp]
  \centering
  \includegraphics[width=0.95\linewidth]{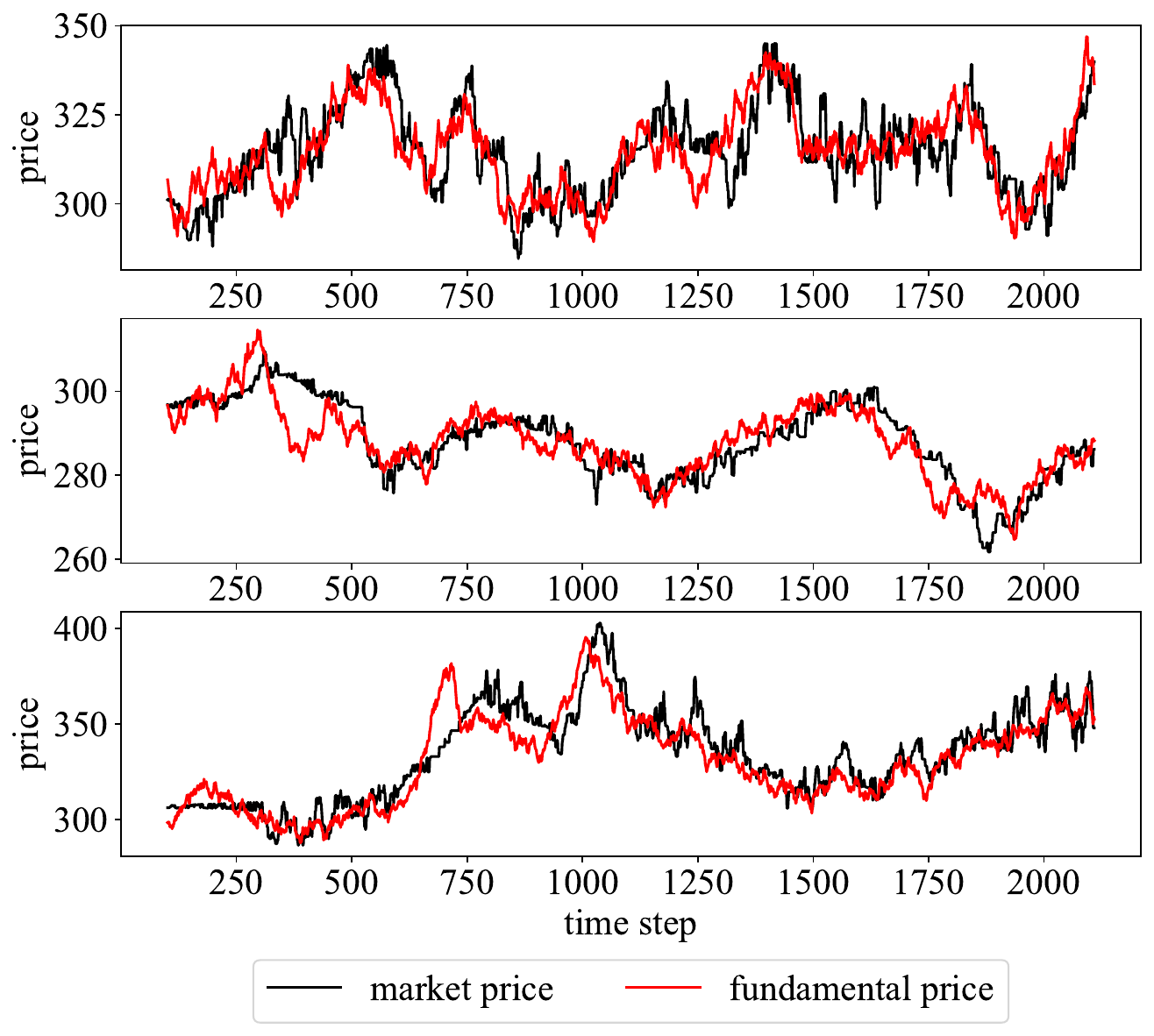}
  \caption{Representative examples of simulated market price series generated by our method.}
  \label{Fig:simulated_prices}
\end{figure}

Figure~\ref{Fig:simulated_prices} shows examples of simulated market prices generated by our method. Market prices generally follow fundamental values, indicating that agents have collectively learned the norm that prices should reflect fundamentals. At the same time, short-term deviations and fluctuations suggest market noise and diverse agent behaviors. Overall, the model strikes a balance between agent-level diversity and coherent market structure. Agents act differently, yet their interactions produce organized price dynamics.

\section{Conclusion}
This study sought a bottom-up understanding of how financial market dynamics emerge by analyzing the roles of heterogeneous preferences and learning mechanism. In our MARL-based ABM, agent preferences are embedded in both observations and rewards, and agents learn a shared-policy that yields behaviors aligned with their preferences.
Experiments showed that (i) agents differentiated their behavior according to preferences within a unified learning framework, leading to niche specialization, and (ii) interactions among differentiated agents reproduced realistic market dynamics. These findings demonstrate emergence from emergence: micro-level differentiation giving rise to macro-level dynamics.
This multilayered emergence parallels biological organization, where cells differentiate into organs and organs self-organize into a body, suggesting that financial markets likewise form higher-order structures through heterogeneous preferences and adaptive learning.

% hierarchical emergence -> Loop of Emergence
Future research may extend such a hierarchical emergence to the {\em loop of emergence} by incorporating micro-macro loops in which agents repeatedly adapt to and shape evolving environments. Such a co-evolutionary dynamic enables the simulation as nonequilibrium systems, capturing more realistic patterns of structural change in financial markets.

% \begin{acks}
% WRITE ME
% \end{acks}

%%%%%%%%%%%%%%%%%%%%%%%%%%%%%%%%%%%%%%%%%%%%%%%%%%%%%%%%%%%%%%%%%%%%%%%%

%%% The next two lines define, first, the bibliography style to be 
%%% applied, and, second, the bibliography file to be used.

\bibliographystyle{ACM-Reference-Format} 
\bibliography{sample}

%%%%%%%%%%%%%%%%%%%%%%%%%%%%%%%%%%%%%%%%%%%%%%%%%%%%%%%%%%%%%%%%%%%%%%%%

\end{document}